\documentclass[journal=accacs,manuscript=article,layout=twocolumn, 9pt]{achemso}
\usepackage[version=3]{mhchem} % Formula subscripts using \ce{}

\newcommand\tabcaption{\def\@captype{table}\caption}
\newcommand\figcaption{\def\@captype{figure}\caption}
\usepackage{amsmath}
\usepackage{chemformula}    % \ch command for chemical formulae
\usepackage{bm}
\usepackage{multirow}
\usepackage{mdframed}
\usepackage{rotating}
\usepackage{booktabs}
\usepackage{etoolbox}
\usepackage{enumitem}
\usepackage{subcaption}
\usepackage[nonumberlist]{glossaries}
\mciteErrorOnUnknownfalse
\usepackage{xcolor}
\usepackage{comment}
\usepackage{array}
\usepackage[hidelinks]{hyperref}
\newcounter{magicrownumbers}
\preto\tabular{\setcounter{magicrownumbers}{0}}
\usepackage{todonotes}
\usepackage{siunitx}

\usepackage{droidsans}
\usepackage{charter}
\usepackage{setspace}
\usepackage[compact]{titlesec}

\newacronym{SHE}{SHE}{standard hydrogen electrode}
\newacronym{pymatgen}{\textit{pymatgen}}{python materials genomics}
\newacronym{GNN}{GNN}{graph neural network}
\newacronym{ASE}{ASE}{Atomic Simulation Environment}
\newacronym{DFT}{DFT}{Density Functional Theory}
\newacronym{HER}{HER}{hydrogen evolution reaction}
\newacronym{OER}{OER}{oxygen evolution reaction}
\newacronym{ORR}{ORR}{oxygen reduction reaction}
\newacronym{MAE}{MAE}{mean absolute error}
\newacronym{OCP}{OCP}{Open Catalyst Project}
\newacronym{OC20}{OC20}{Open Catalyst 2020}
\newacronym{OC22}{OC22}{Open Catalyst 2022}
\newacronym{PCA}{PCA}{principal component analysis}
\newacronym{VASP}{VASP}{Vienna \textit{Ab Initio} Simulation Package}
\newacronym{WNA}{WNA}{water nucleophilic attack}
\newacronym{I2M}{I2M}{oxo-coupling mechanism}
\newacronym{LOER}{LOER}{Lattice oxygen evolution reaction}
\newacronym{MvK}{MvK}{Mars van Krevelen}
\newacronym{ML}{ML}{Machine learning}
\newacronym{ClER}{ClER}{chlorine evolution reaction}
\newacronym{S2EF}{S2EF}{\textit{\textbf{Structure to Energy and Forces}}}
\newacronym{PAW}{PAW}{Projected Augmented Wave}
\newacronym{PBE}{PBE}{Perdew-Berke-Ernzerhof}
\newacronym{GGA}{GGA}{generalized gradient approximation}
\newacronym{SA:V}{SA:V}{surface area-to-volume ratio}
\newacronym{MMI}{MMI}{maximum Miller index}
\newacronym{PDS}{PDS}{potential determining step}
\glsaddall

\newcommand{\chemeng}{ William A. Brookshire Department of Chemical and Biomolecular Engineering and Texas Center for Superconductivity (TcSUH), University of Houston, 4226 Martin Luther King Boulevard, Houston, TX 77204, USA.}
\newcommand{\coa}{Indicates equal contributions}
\newcommand{\ece}{Department of Electrical and Computer Engineering, University of Houston, 4226 Martin Luther King Boulevard, Houston, TX 77204, USA.}
\newcommand{\shell}{Shell Global Solutions, Houston, TX 77082, USA.}
\newcommand{\ist}{Department of Information Science Technology, University of Houston, 14004 University Boulevard \#318, Sugar Land, TX 77479, USA.}
\newcommand{\etca}{Energy Transition Campus Amsterdam, Shell Global Solutions International B.V. Grasweg 31, 1031 HW Amsterdam, the Netherlands.}
\newcommand{\bang}{Shell Technical Center Bangalore, 7, Bengaluru Hardware Park KIADB Industrial Park North, Mahadeva Kodigehalli, Bengaluru, Bangalore, Karnataka 562149, India.}

\newcommand\nmaterials{4,119}
\newcommand\npredictions{6,068,572}

\title[Short title]{Rational design of nanoscale stabilized oxide catalysts for OER with OC22}

\author{Richard Tran}
\affiliation{\coa}
\alsoaffiliation{\chemeng}
\email{rtran17@uh.edu}
\author{Liqiang Huang}
\affiliation{\coa}
\alsoaffiliation{\ece}
\author{Yuan Zi}
\affiliation{\ece}
\author{Shengguang Wang}
\affiliation{\chemeng}
\author{Benjamin M. Comer}
\affiliation{\shell}
\author{Xuqing Wu}
\affiliation{\ist}
\author{Stefan J. Raaijman}
\affiliation{\etca}
\author{Nishant K. Sinha}
\affiliation{\bang}
\author{Sajanikumari Sadasivan}
\affiliation{\bang}
\author{Shibin Thundiyil}
\affiliation{\bang}
\author{Kuldeep B. Mamtani}
\affiliation{\bang}
\author{Ganesh Iyer}
\affiliation{\bang}
\author{Lars Grabow}
\alsoaffiliation{\chemeng}
\author{Ligang Lu}
\affiliation{\shell}
\email{ligang.lu@shell.com}
\author{Jiefu Chen}
\affiliation{\ece}
\email{jchen84@uh.edu}

\keywords{Catalysis, oxides, renewable energy, datasets, machine learning, high-throughput screening, nanoscale stability}

\let\oldmaketitle\maketitle
\let\maketitle\relax

\begin{document}

\twocolumn[
\begin{@twocolumnfalse}
\oldmaketitle
\begin{abstract}
The efficiency of \ce{H2} production via water electrolysis is typically limited to the sluggish oxygen evolution reaction (OER). As such, significant emphasis has been placed upon improving the rate of OER through the anode catalyst. More recently, the Open Catalyst 2022 (OC22) framework has provided a large dataset of density functional theory (DFT) calculations for OER intermediates on the surfaces of oxides. When coupled with state-of-the-art graph neural network models, total energy predictions can be achieved with a mean absolute error as low as 0.22 eV. In this work, we interpolated a database of the total energy predictions for all slabs and OER surface intermediates for \nmaterials~oxide materials in the original OC22 dataset using pre-trained models from the OC22 framework. This database includes all terminations of all facets up to a maximum Miller index of 1\color{black}. To demonstrate the full utility of this database, we constructed a flexible screening framework to identify viable candidate anode catalysts for OER under  varying reaction conditions for bulk, surface, and nanoscale Pourbaix stability as well as material cost, overpotential, and metastability\color{black}. From our assessment, we were able to identify 122 and 68 viable candidates for OER under the bulk and nanoscale regime respectively\color{black}. 
\end{abstract}

\end{@twocolumnfalse}
]

\clearpage
\section{Introduction} 
As the global focus shifts increasingly towards renewable energy, there has been a significant rise in the demand for cost-effective and environmentally sustainable energy storage and transmission methods.\color{black}~Electrochemical water splitting, or water electrolysis, is a sustainable and promising means of evolving\color{black}~\ce{H2} thanks to the wide abundance of water. This process involves two coupled half-reactions: \gls{HER} and the significantly slower \gls{OER} which has primarily been the bottleneck in advancing water-splitting technology. The search for a highly active anode catalyst for \gls{OER} is therefore paramount to the realization of practical water splitting technology\cite{Jamesh2018, Yuan2020, Kim2018, Suen2017}. 

Transition metal oxides are a promising class of catalysts for \gls{OER} due to their varying oxidation states for more efficient multi-electron transfer, stability under highly acidic conditions favorable towards \gls{OER}, and active undercoordinated transition metals sites. In regard to commercialized catalysts, \ce{IrO2} and \ce{RuO2} are the benchmark catalysts for \gls{OER}, exhibiting low overpotentials (an indicator of activity) of 0.25 to 0.5 V under acidic conditions\cite{Lee2012}. However, the material cost of precious metals (\$18,315 and \$155,727 per kg for \ce{RuO2} and \ce{IrO2} respectively as of March 2021\cite{dailymetalprice, apmex, metalary} with a price variation of $\pm$ \$9,969 for \ce{RuO2} and $+$ \$80,370 for \ce{IrO2}\color{black}~in the last 24 years) limits their widespread adoption. Consequently, there is much desire to identify cheaper materials for catalysts in \gls{OER} while maintaining similar performance. 

Despite the abundance of unary and binary oxides, very few are capable of exhibiting both high catalytic activity and stability under operating conditions. Computational analysis performed by \citet{Wang2020} indicated only 68 bimetallic oxides from a pool of 47,814 were stable with a Pourbaix decomposition energy ($\Delta G_{PBX}$) of 0.5 eV/atom or less under acidic conditions (pH=0) and an applied potential cycle between 1.2 and 2.0 V. \citet{Gunasooriya2020} calculated the overpotential of these candidate oxides and identified only 11 nonbinary metal oxides with a promising overpotential of less than 0.85 V.

It is possible that many materials filtered out by the highly discriminant electrochemical stability criteria can also exhibit competitive overpotentials if stabilized. Nanoscale stability, elemental doping, and the introduction of oxygen vacancies have been demonstrated to be effective means of improving stability\cite{Jamesh2018, abed2023multi, Fu2021}. Nanoscaling in particular presents a promising avenue for stabilizing oxides under operating conditions while exposing a greater number of active sites through the increasing \gls{SA:V}\cite{Kang2014, Sun2016b, Navrotsky2011}. %At the nanoscale, the properties of the surface, as oppose to the bulk, will begin to dictate the overall properties of the nanoparticle due to the rapid increase in the \gls{SA:V}. Consequently, the relative thermodynamic stability of materials in the same chemical systems can result in unstable compounds becoming stable as the lower surface energies of these compounds begin influence the formation energy at the nanoscale.

However, an accurate evaluation of the nanoparticle formation energy not only requires the thermodynamic contributions of the bulk, but the surface as well. Doing so requires an ensemble of expensive \gls{DFT} calculations of the bare surfaces for one material. Likewise, accurate evaluations of the overpotential requires an even larger set of calculations for all the surface intermediates participating in \gls{OER}. As such, doing so for the massive pool of binary and unary oxide materials available will quickly become computationally expensive and unfeasible. 

\gls{ML} potentials and screening frameworks have recently contributed significantly in rapidly predicting candidate catalysts without the need for expensive systematic \gls{DFT} screening \cite{Li2020, Rao2020, Abdelfatah2019, Malek2021, Back2019}. Among these efforts, the \gls{OCP}\cite{Chanussot2021} framework has stood out as having the largest dataset of carefully curated \gls{DFT} calculations for non-oxide slabs and surface intermediates to-date. \gls{ML} models pre-trained with the \gls{OCP} has allowed for large scale interpolation efforts to predict binding energies, enabling high-throughput screening efforts to identify viable catalyst candidates\cite{Tran2022} and explore fundamental surface chemistry\cite{Price2022}. More recently, the \gls{OC22} framework has expanded upon this dataset by incorporating random combinations of 4,728 oxide materials, Miller indices up to 3, and surface intermediates involved in \gls{OER} \cite{Tran2022b}. These efforts have yielded predictive \gls{ML} models with total energy \gls{MAE}s of less than 0.22 and 0.69 eV for \textit{in domain} (materials observed during training) and \textit{out of domain} predictions, respectively. 

In this manuscript, we utilized a pre-trained model from the \gls{OC22} framework to interpolate the surface energies and \gls{OER} binding energies for \nmaterials~\textit{in domain} oxide materials on all facets up to a \gls{MMI}\color{black}~of 1. To demonstrate the applicability of our interpolated database, we constructed a high-throughput screening framework with a set of progressive criteria that can be modified or expanded upon for ease of customizeability in order to evaluate the commercial and practical viability of each material for \gls{OER}. Our general framework evaluates materials based on thermodynamic stability, overpotential, and material cost. We also expand upon other screening criteria such as the possibility of nanoscale stabilization or the faceting of surfaces on the equilibrium crystal structure. We propose 190\color{black}~possible candidates for \gls{OER} under the bulk and nanoscale regime respectively that warrant further experimental investigation.

\section{Methods} 
All analysis were performed using the \gls{pymatgen}\cite{Ong2013, Persson2012, Singh2017, Patel2019} and \gls{ASE}\cite{HjorthLarsen2017} packages.

\subsection{Slab generation}

We described all facets up to a \gls{MMI} of 1 containing an atomic and vacuum layer of 12.5 \AA~thick. The bulk materials used for slab construction in this study were obtained from the Materials Project\cite{Jain2013}. We also considered all terminations for each facet (\textit{sans} slabs exceeding 200 atoms) while maintaining equivalent surfaces on both sides of the slab which consequently resulted in non-stoichiometric slabs with respect to the bulk formula. Although the original \gls{OC22} dataset covered 4,732 distinct bulk oxide materials, the conventional unit cell of some of these materials contains over 100 atoms making the construction of slabs exceed our 200 atom limit for the majority of facets considered. As such, we limit our study to slabs constructed from unit cells of less than 100 atoms. Furthermore, slabs constructed from a select number of materials resulted in the forces in the \gls{ML} model being unconverged. Thus, we ommitted 609 from the original 4,732 materials in the \gls{OC22} dataset that exhibited these behaviors, with our final bulk set containing 3,823 binary (A-B-O) and 296 unary (A-O) oxides. For all slabs constructed, we modelled the surface intermediates of \ce{O^*}, \ce{OH^*}, and \ce{OOH^*}. To avoid periodic interactions between the adsorbates, all slabs were expanded along the length and width to at least 8 \AA. We assumed all adsorbates bind through the \ce{O} atom on available undercoordinated metal sites. All bare surface and surface intermediate models were constructed using the python framework adapted from~\gls{OC22}\cite{Tran2022b, Sun2013a, Montoya2017}\color{black}.

\subsection{DFT and machine learning settings}

All \gls{DFT} calculations were performed using the \gls{VASP}\cite{Kresse1993, Kresse1994, Kresse1996a, KresseG1996b} within the \gls{PAW}\cite{Blochl1994} approach. We modeled the exchange-correlation effects with the \gls{PBE} \gls{GGA} functional\cite{Perdew1996}. All calculations were performed with spin-polarization with a plane wave energy cut-off of 500 eV. The energies and atomic forces of all calculations were converged to within $1 \times 10^{-4}$ eV and 0.05 eV \AA$^{-1}$, respectively. We used $\Gamma$-centered k-point meshes of $\frac{30}{\textit{a}} \times \frac{30}{\textit{b}} \times 1$ for slab calculations, with non-integer values rounded up to the nearest integer. We also apply a Hubbard U correction to chemical systems as suggested by the Materials Project\cite{Jain2011g} to account for missing electron interactions.

We used a pre-trained model for the \gls{S2EF}-\textit{Total} task from the \gls{OC22} framework to perform all machine learning predictions of the relaxed structure and total energy. The entirety of the \gls{OC20} dataset (1,281,040 \gls{DFT} relaxations) was used to train an \gls{S2EF}-\textit{Total} model which was subsequently fine-tuned with the \gls{OC22} dataset (62,331 \gls{DFT} relaxations) to better predict the total energies of oxide surfaces and surface intermediates. The model was trained using the GemNet-OC architecture\cite{Gasteiger2022} due to its superior performance in energy predictions when compared to other \gls{GNN} architectures as a consequence of its improved capturing of long-range and quadruplet interactions. 

For further details regarding additional parameters used in \gls{VASP} or the construction of the machine learning model, we refer the reader to~\citet{Tran2022b}.

All \gls{OC22} \gls{S2EF} predictions and \gls{DFT} calculations of slabs were performed with selective dynamics. In regards to surface energy predictions, both the bottom most and topmost layer of atoms within 1.25~\AA~were allowed to relax in the bare slab in order to ensure both surfaces had equal surface energy contributions. For the adsorption energies, only the topmost layer of atoms within 1.25~\AA~and any adsorbates were allowed to relax in the bare slab and surface intermediates. To avoid inadvertent desorption and dissociation of adsorbates during our \gls{ML} relaxation, we applied a spring constant of 7.5 eV~\AA$^{-2}$ between all adsorbate atoms to preserve the identity of the molecule and between the adsorbate and host surface atoms whenever the adsorbate drifts 2~\AA~away from its initial position along the axis perpendicular to the surface\cite{Peterson2014, Musielewicz2022}. Similarly, we also applied the same restorative force to all surface atoms when the relaxed trajectory exceeds 1~\AA~from the initial position of the ions to avoid drastic surface reconstruction.  While this approach yields better interpretability for relaxed adsorption geometries, we recognize the inherent artificiality of these constraints and acknowledge the potential for desorption and dissociation. Consequently, we further implemented \gls{ML} relaxation without such constraints on relaxed geometries exhibiting low overpotentials to verify the absence of desorption and dissociation phenomena. All values of overpotentials and Gibbs free energies reported in the main manuscript as well as Tables S1-S14 are relaxed without these additional constraints unless stated otherwise. Comparisons of the overpotential and Gibbs free energy obtained with and without these constraints can be found in the ESI.
\subsection{Surface thermodynamics}

\begin{figure}[ht]
    \centering
    \includegraphics[width=1\linewidth]{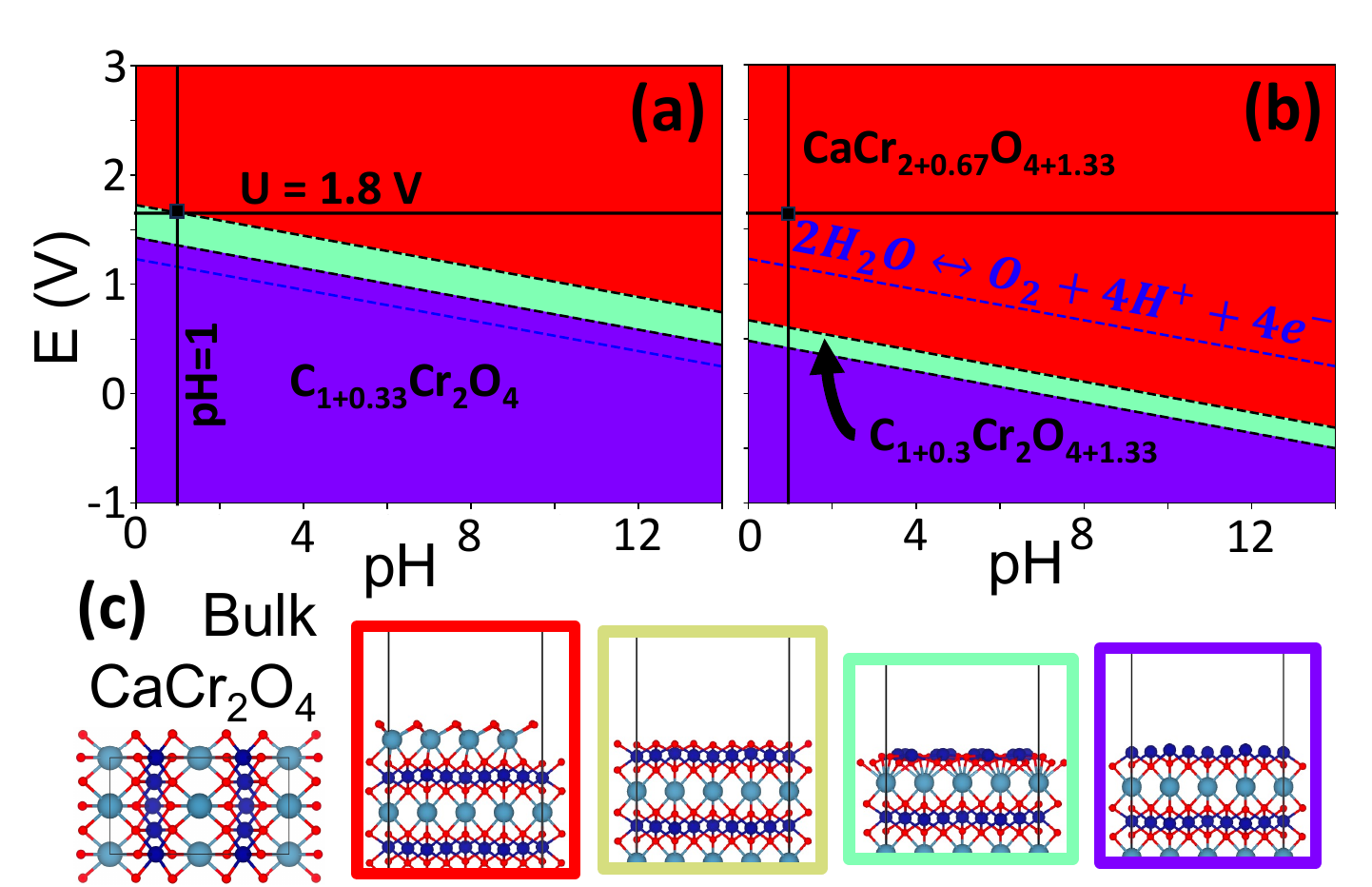}
    \caption{Surface Pourbaix diagram of the (001) facet of \ce{CaCr2O4} at $T=80 ^oC$ under $\Delta \mu_{Cr}=-4$ eV (a) and $\Delta \mu_{Cr}=0$ eV (b). The blue dashed line indicates the equilibrium conditions for \gls{OER} while the black box indicates the desired conditions for \gls{OER} ($pH=1$ and $U = 1.8$ V). The phases in (a) and (b) are color coded with the corresponding terminations of the (001) facet in (c).}
    \label{fig:pourbaix}
\end{figure}

All bare slabs of formula \ce{A_{nx}B_{ny+k}O_{nz+j}}, with A and B being two metal components, are constructed from a bulk ternary oxide of \ce{A_{x}B_{y}O_{z}}. We can calculate the surface energy of any slab of \ce{A_{nx}B_{ny+k}O_{nz+j}} with the following:
\begin{equation}
\label{eq:surface_energy_chempot}
\begin{split}
    & \gamma(\mu_B, \mu_{\ce{O2}}) = \\
    & \frac{E_{slab}^{\ce{A_{nx}B_{ny+k}O_{nz+j}}} - (nE^{\ce{A_xB_yO_z}}_{bulk} + k\mu_B + j\frac{1}{2}\mu_{O_2})}{2A}
\end{split}
\end{equation}
where $E_{slab}^{\ce{A_{nx}B_{ny+k}O_{nz+j}}}$ is the total energy of the bare slab, $E^{\ce{A_xB_yO_z}}_{bulk}$ is the total energy per formula unit of the bulk crystal, and $A$ is the surface area. We used correction values from the Materials Project when evaluating $\gamma$ to account for the mixing of quantities determined with GGA and GGA+U\cite{Persson2012, Jain2011g, Wang2021}. The chemical potentials, $\mu_i$, accounts for any nonstoichiometric species (with respect to bulk stoichiometry) in the slab formula.

The chemical potential of oxygen ($\Delta \mu_{\ce{O2}}$) can be referenced to the electrochemical decomposition of water to $\ce{O2_{(g)}}$:
\begin{equation}
\label{eq:water_breakdown}
2\ce{H2O_{(g)}} \rightarrow 4(\ce{H^+} + e^-) + O_{2(g)}: 4.92~eV
\end{equation}
which allows us to rewrite $\mu_{\ce{O2}}$ (and thereby $\gamma$) as a function of $pH$ and applied potential ($U$) as such:
\begin{equation}
\begin{split}
\label{eq:muO_pH_U}
    & \Delta \mu_{O_2} = \\
    & 4.92 + 2\mu^o_{\ce{H2O}} - \\
    & 4(\frac{1}{2}\mu^o_{\ce{H2}} - eU - k_BTpHln10) + \Delta G^{O^*}_{corr}
\end{split}
\end{equation}
where $\Delta G^{O^*}_{corr}$ corrects for the Gibbs free energy of excess or deficient oxygen at the surface (see~\citet{Tran2022b} and~\citet{Gunasooriya2020} for details). We will assume typically employed operating conditions for acidic \gls{OER} ($pH = 1$ and $U = 1.8$ V at $T = $ 80 $^oC$ or 60 $^oC$)\cite{Wang2020, Shi2022}\color{black}~($\Delta \mu_{O_2} = -1.30$ eV) when assessing the surface energy of all materials as illustrated in Figure~\ref{fig:pourbaix}(a and b). For slabs containing excess or deficient metal (B) species, the chemical potential of component \ce{B} is conventionally referenced with respect to the per atom energy of the ground state bulk crystal of pure component B (e.g. $\mu_{Fe} = \Delta \mu_{Fe}+E^{DFT}_{BCC, Fe}$). By varying the chemical potential of component \ce{B}, we can stabilize different surface terminations of the same facet as shown in the surface Pourbaix diagram in Figure~\ref{fig:pourbaix}(a and b).

\begin{figure}[ht]
    \centering
    \includegraphics[width=1\linewidth]{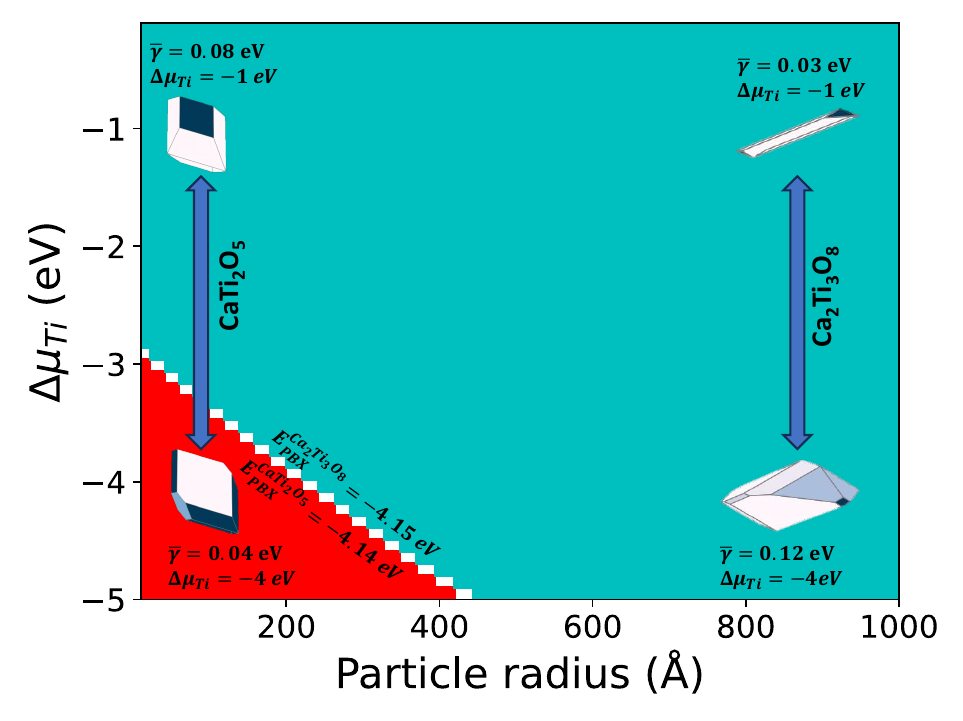}
    \caption{Nanoscale stability phase diagram for the Ca-Ti-O chemical system plotted using $G_f^{NP}$ as a function of the chemical potential of \ce{Ti} ($\Delta \mu_{Ti}$) and nanoparticle radius. The red region represents the less stable (higher $E_{PBX}$) phase (\ce{CaTi2O5}) while the cyan region represents the ground state phase (\ce{Ca2Ti3O8}). The nanoparticle morphology and surface energy change as a function of $\Delta \mu_{Ti}$ indicated by the inset Wulff shapes.}
    \label{fig:nanoscale_stability}
\end{figure}

To determine the nanoscale stability of metastable and unstable materials under operating conditions, we assessed the nanoparticle formation energy given by:

\begin{equation}
\begin{split}
\label{eq:G_np_form}
    & G_f^{NP} = \\
    & E_V(pH, V, T)(\frac{4}{3}\pi r^3) + \bar{\gamma}(pH, V, T, \Delta \mu_{M})(4\pi r^2)
\end{split}
\end{equation}
whereby $E_V$ is the Pourbaix formation energy per volume of the unit cell, $\bar{\gamma}$ is the weighted surface energy of the Wulff shape (an analogue to the nanoparticle morphology), and $r$ is the radius of the nanoparticle. Detailed explanation of these quantities can be found in the ESI. Figure~\ref{fig:nanoscale_stability} demonstrates how a less stable compound (\ce{CaTi2O5}) can become more stable than the ground state compound as nanoparticle size and $\Delta \mu_{Ti}$ decreases. The size effect will change the relative contribution of surface energy and bulk Pourbaix formation energy to $G_f^{NP}$ while $\Delta \mu_{Ti}$ changes the overall particle morphology and thereby surface energy of the particle. 

\begin{figure*}
    \centering
    \includegraphics[width=1\linewidth]{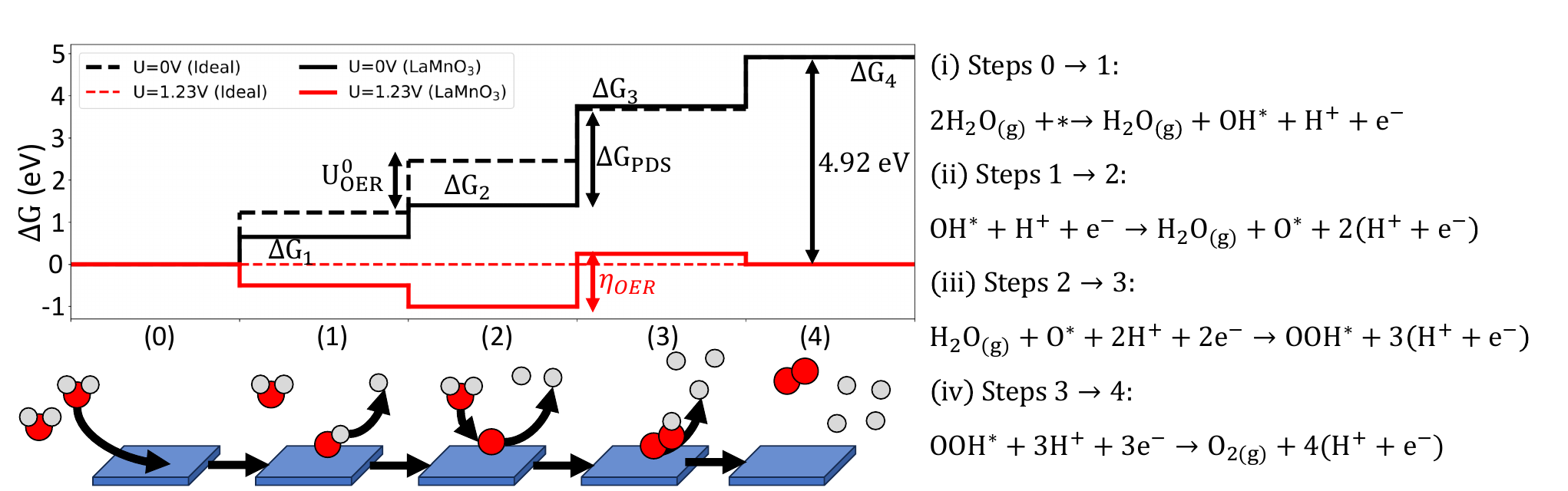}
    \caption{Reaction diagram for the \gls{WNA} on an ideal catalyst (dashed lines) and \ce{LaMnO3} (100) surface (solid lines) at 0 V (black) and the equilibrium potential of \gls{OER} (red). An illustration of each reaction step is shown at the bottom (red circles are oxygen while grey circles are hydrogen/protons). Reaction energies for \ce{LaMnO3} were derived from~\citet{Jaramillo2011}. The chemical equation between each reaction step (1-4) is listed on the right (\textit{i-iv}).}
    \label{fig:rxn_pathway}
\end{figure*}

The scope of this study will explore the \gls{WNA} mechanism for \gls{OER}, a four step mechanism where two water molecules sequentially bind to a metal at the surface and release an electron-proton pair at each step as shown in Figure~\ref{fig:rxn_pathway}\cite{Jaramillo2011}. We realize and emphasize that the~\gls{WNA} mechanism is one of many approximations for modeling \gls{OER} and that it is possible for certain materials to prefer alternative mechanisms e.g. the oxo-coupling mechanism or lattice oxygen evolution reaction\cite{Fabbri2014, Reier2017, Dau2010, Fabbri2018, Gonzalez2021}. However, we focus on the \gls{WNA} mechanism on account of its ubiquity in computational studies\cite{Zhang2020, Fornaciari2022, Huang2021, Nishimoto2020, Mefford2020} and supporting experimental evidence\cite{Naito2021, Reier2017}.\color{black}~We can determine the overpotential for this reaction by identifying the largest energy difference between each step, reaction energy ($\Delta G_{rxn}$) with the following:
\begin{equation}
\begin{split}
\label{eq:overpotential}
    & \eta = \\
    & max(\Delta G^{i}, \Delta G^{ii} - \\
    & \Delta G^{i}, \Delta G^{iii} - \Delta G^{ii}, 4.92 - \Delta G^{iii})/e - 1.23 V
\end{split}
\end{equation}
where $\Delta G^{i}$, $\Delta G^{ii}$, and $\Delta G^{iii}$ are the Gibbs free energy of each reaction step, 4.92 eV is the Gibbs free energy to dissociate two water molecules into \ce{O_2} and 4(\ce{H^+}+\ce{e^-}) shown in Equation~\ref{eq:water_breakdown}, and 1.23 V is the equilibrium potential for water decomposition. Here, the step corresponding to the largest value of $\Delta G_{rxn}$ is also called the \gls{PDS}\color{black}. We can derive the Gibbs free energy of each step listed in Equations $i-iv$ (see Figure~\ref{fig:rxn_pathway}) as such:
\begin{equation}
\label{eq:gads1_corr}
\Delta G^{i} = E_{ads}^{\ce{OH^*}} + \Delta G^{OH^*}_{corr} + \mu_{H^+} + \mu_{e^-}
\end{equation}
\begin{equation}
\label{eq:gads2_corr}
\Delta G^{ii} = E^{\ce{O^*}} + \Delta G^{O^*}_{corr} + 2(\mu_{H^+} + \mu_{e^-})
\end{equation}
\begin{equation}
\label{eq:gads3_corr}
\Delta G^{iii} = E^{\ce{OOH^*}} + \Delta G^{OOH^*}_{corr} + 3(\mu_{H^+} + \mu_{e^-})
\end{equation}
where $E_{ads}^{\ce{OH^*}}$, $E_{ads}^{\ce{O^*}}$, and $E_{ads}^{\ce{OOH^*}}$ are the electronic adsorption energies of the intermediates for \gls{OER} and $G^{OH^*}_{corr}$, $G^{O^*}_{corr}$, and $G^{OOH^*}_{corr}$ are correction terms for the Gibbs free energy derived in the ESI of \gls{OC22}\cite{Tran2022b}. 

To minimize the number of predictions needed, we will begin by using a quick scaling relationship given by $\Delta G^{iii} = \Delta G^{i} + 3.26$\cite{Gunasooriya2020, Tran2022b} to estimate $\Delta G^{iii}$. This approach is particularly beneficial for \ce{OOH^*}, where the significantly greater rotational freedom leads to a substantial increase in potential adsorbate placements. We will then perform additional predictions for $E^{\ce{OOH^*}}$ for surfaces exhibiting promising activity ($\eta < 0.75$ V) using Equation~\ref{eq:gads3_corr} to more accurately determine overpotential. \gls{ML} relaxations exhibiting dissociation or desorption of intermediates are omitted in any interpretation of $\eta$. More details regarding dissociation and desorption events occurring in the dataset as well as a comparison between $\eta$ obtained with Equation~\ref{eq:gads3_corr} and scaling relationships can be found in the ESI.

Only the most stable site for \ce{OH^*} dictated by $E^{\ce{OH^*}}_{ads}$ from a set of considered adsorption sites on the same surface is considered when determining $\eta$. We maintain the adsorption site corresponding to the most stable site of \ce{OH^*} when considering $E_{ads}^{\ce{O^*}}$ and $E_{ads}^{\ce{OOH^*}}$.

Further details regarding the derivation of all thermodynamic quantities and scaling relationships can be found in the ESI.

\section{Results and Discussion} 
\subsection{Database scope and usage}

\begin{figure}[ht]
    \centering
    \includegraphics[width=1\linewidth]{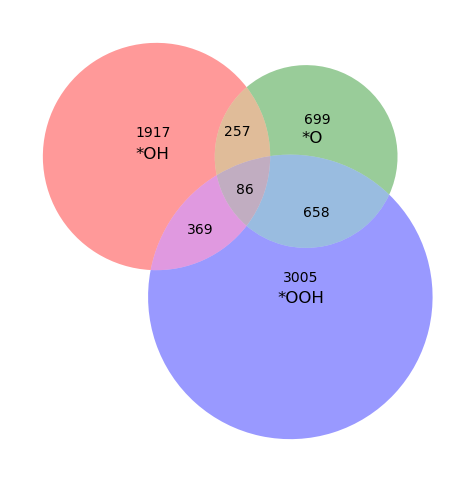}
    \caption{Scope of surface intermediates for \gls{OER} calculated using \gls{DFT} in the \gls{OC22} dataset. Overlaps indicate the number of surfaces where different intermediates (\ce{OH^*}, \ce{O^*}, or \ce{OOH^*}) are calculated for the same surface.}
    \label{fig:oc22_sample}
\end{figure}

We emphasize that the purpose of both the \gls{OC20} and \gls{OC22} datasets was to establish a large and diverse set of DFT calculated surfaces and surface intermediates for the purpose of training \gls{ML} potentials generalized to infer the total \gls{DFT} of any slab and adsorbate combination. To maximize the diversity of the sample set, the \gls{OCP} curated \gls{DFT} calculations of randomly selected combinations of materials, surfaces, and adsorbates. The scope of the dataset \textbf{does not} encompass a comprehensive database for evaluating $\eta$ directly, as many necessary data points are missing illustrated in Figure~\ref{fig:oc22_sample}. The construction of a comprehensive database is a herculean task that is immeasurably costly and time consuming with \gls{DFT} alone. However, by consolidating the \gls{S2EF}-\textit{total} model to predict thermodynamic overpotentials, we can potentially infer the catalytic activity of large material datasets for \gls{OER} within a reasonable degree of error with respect to \gls{DFT}.

Using the previously developed \gls{ML} models, we systematically extrapolated the total \gls{DFT} energy of all terminations for all facets up to a \gls{MMI}=1\color{black}~for all \nmaterials~materials considered in this study. We emphasize that although the original training pool does consider selections of bare and adsorbed surfaces up to \gls{MMI}=3, the complete dataset is biased towards facets with a \gls{MMI}=1 with 39,573, 15,482, and 7,276 data points of \gls{MMI}=1, 2 and 3 respectively. We expect this bias to allow for better predictions of the facets (all of which exhibit \gls{MMI}=1) considered in this work\color{black}. We then extrapolated the total energy of all metal adsorption sites on all surfaces for \ce{O^*} and \ce{OH^*} (and \ce{OOH^*} when $\eta < 0.75$ V)\color{black}. Table~\ref{tab:table_scope} summarizes the scope of our extrapolated database. In contrast to our interpolation efforts, the size of the \gls{OC22} dataset for \ce{O^*} and \ce{OH^*} is only 0.1\% of the\color{black}~predicted dataset in this work. Despite requiring orders of magnitude less computational resources than a \gls{DFT} dataset of the same size, the estimated cost to produce our dataset is still 9,473.9 GPU-hrs. (an average rate of one prediction every 12.5 seconds), a significant amount of resources. By making this database freely available to the scientific community, users interested in performing similar high-throughput screening exercise or fundamental analysis can do so without the enormous cost in GPUs. The entire database including the initial and relaxed structures and total energies can be accessed through the University of Houston Dataverse Repository\cite{APJFTM_2023}. Details regarding the database metadata are given in the ESI.

For the 12,922 surfaces exhibiting $\eta < 0.75$ V, we performed an additional \gls{ML} relaxation step without the application of constraints on surface relaxation, adsorbate dissociation, and desorption. While only 7\% of \ce{OH^*} and 1\% of \ce{O^*} intermediates exhibited dissociation and desorption event, an overwhelming number of events (50\%) were associated with the \ce{OOH^*} intermediates. Consequently, predictions of $\eta$ exhibiting these events could not be interpreted and were ignored in our final assessment of overpotential. However, we emphasize that the occurrence of dissociation and desorption does not disqualify the possibility of these surfaces being catalytically active in alternative reaction mechanisms such as the oxo-coupling mechanism or lattice oxygen evolution reaction\cite{Fabbri2014, Reier2017, Dau2010, Fabbri2018, Gonzalez2021}.\color{black}

\begin{table}[ht]
    \centering
\scalebox{0.8}{
\begin{tabular}{cccc}
    \hline
    \multicolumn{4}{c}{\textbf{Predictions}: \npredictions} \\
    \hline
    \multicolumn{4}{c}{\textbf{Materials}: \nmaterials} \\
    \hline
    \multicolumn{4}{c}{\textbf{Ave. \# slabs per material}: 47} \\
    \hline
    \textbf{\ce{OH^*}} & \textbf{\ce{O^*}} & \textbf{\ce{OOH^*}} & \textbf{*}\\
    1,972,166 & 667,266 & 3,237,238 & 191,902\\
    \hline
    \hline
    \multicolumn{4}{c}{\textbf{Predictions w/o spring constraints}}\\
    \hline
    \hline
    \multicolumn{4}{c}{\textbf{Predictions of $\eta$: 12,922}} \\
    \hline
    \multicolumn{4}{c}{\textbf{Dissociation/desorption events}} \\
    \textbf{\ce{OH^*}} & \textbf{\ce{O^*}} & \textbf{\ce{OOH^*}}\\
    908 & 131 & 6,296\\
    \hline
    \multicolumn{4}{c}{\textbf{Predictions of $\eta$ w/o diss./des. events: 6,468}} \\
    \hline
\color{black}

\end{tabular}
    \caption{{Summary of database scope}}
    \label{tab:table_scope}
}
\end{table}
\subsection{High-throughput screening}

\begin{figure}
    \centering
    \includegraphics[width=1\linewidth]{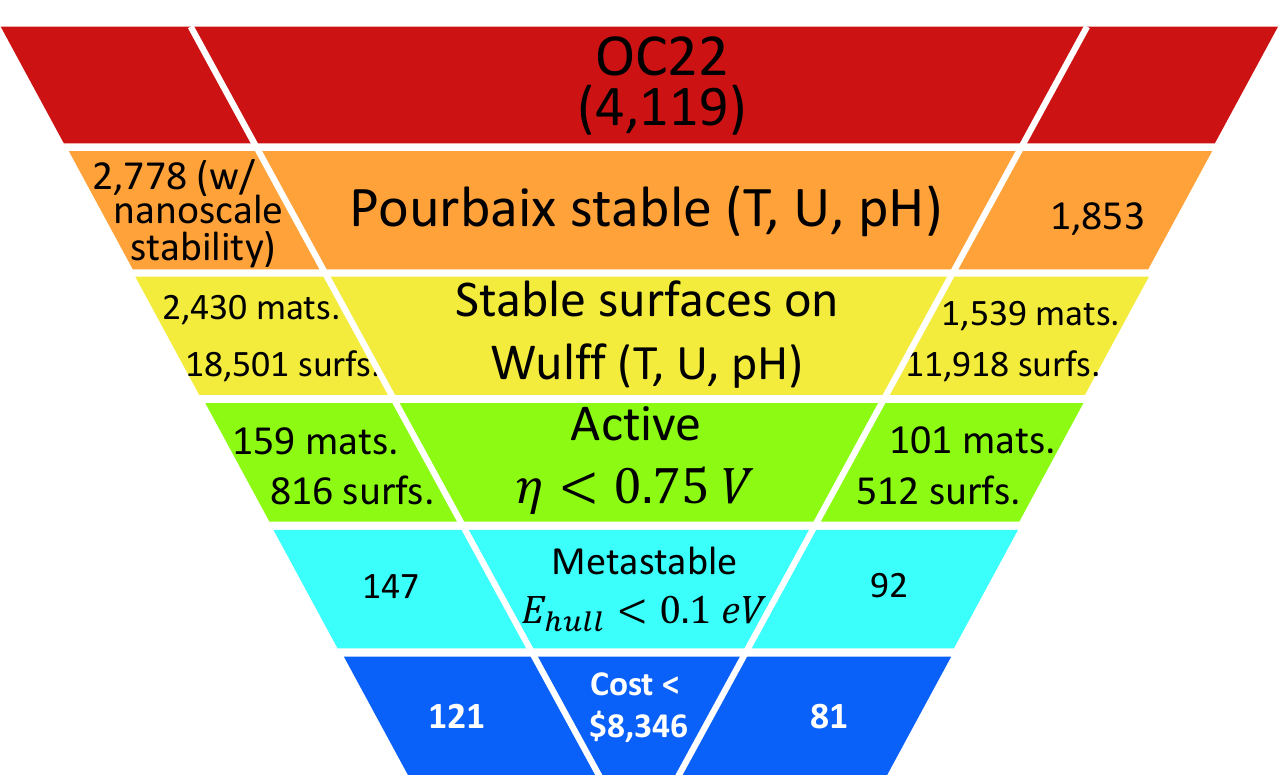}
    \caption{Summary of the screening criteria considered in our high-throughput screening framework. The possible number of candidates that have satisfied each tier is listed on the right for bulk Pourbaix stable materials and on the left when nanoscale stability is possible. The second and third criteria can be further modified by changing the environmental parameters (T, U, pH).}
    \label{fig:screening}
\end{figure}

Figure~\ref{fig:screening}~summarizes the selection criteria that we employed to screen for candidate electrocatalysts for \gls{OER}. The first criterion in our high-throughput screening framework was to determine if a material exists in the original training dataset of \gls{OC22}. As previously demonstrated by~\citet{Tran2022b}, the \gls{OC22} framework is capable of predicting the total \gls{DFT} energies of slabs and surface intermediates of oxides within a mean absolute error of 0.22 eV for 4,119 materials that have been observed in the training dataset. Our database will only interpolate the energies of slabs and surface intermediates amongst these materials (for more information, the reader is redirected to~\cite{Tran2022b}). 

% \textcolor{blue}{In our extensive study of adsorption sites and terminations, we investigated all possible surfaces up to a maximum Miller index of one. This approach led to a sample size of approximately 7,000,000 distinct adsorbate-slab structures. Capitalizing on the modern GPUs, we adopted a parallel computational strategy wherein each GPU processed 8 samples simultaneously. This method enabled us to optimize each sample from its initial structure to its relaxed state, and compute its associated energy in a remarkably short span of 5-15 seconds per sample. Distributing the computations across multiple GPU tasks, we successfully completed the calculations for all seven million samples in a period of 2 months. The speed and efficiency of our method, enhanced by high-speed machine learning predictions, allowed a comprehensive investigation of the surfaces — a feat unachievable with DFT calculations given current computational resources. Additionally, our approach presents the opportunity to identify rare adsorbate-slab structures that could be inadvertently missed due to human biases. After computing these structures, we subjected the samples to a series of screening processes based on specific criteria to identify potential candidates for OER.} 

The second criterion describes the Pourbaix stability, i.e., the electrochemical stability of a material in an aqueous environment. Here, we can interpret Pourbaix stability under a bulk regime (right side of Figure~\ref{fig:screening}). We quantify the Pourbaix stability of the bulk using the Pourbaix decomposition energy ($E_{PBX}$), which is a function of the temperature (T) applied potential (U) and pH of the environment (at T = 80$^o$C, U = 1.8 V, and pH =1)\cite{Patel2019, Singh2017}. Materials with $E_{PBX} = 0$ eV per atom are stable under such conditions while materials with $E_{PBX} >$ 0 eV atom$^{-1}$ are metastable with the likelihood of corrosion increasing with $E_{PBX}$. It was shown experimentally that metastable materials with $E_{PBX} \leq 0.2$ eV per atom are less likely to dissolve or corrode\cite{Jain2019}. However, materials with $E_{PBX}$ as high as 0.5 eV atom$^{-1}$ have also been shown to be stable, albeit with a degree of surface passivation which can inhibit catalytic activity\cite{Singh2017}. We allow any material with $E_{PBX} \leq 0.5$ eV per atom under the aforementioned conditions (see Figure~\ref{fig:pourbaix})(a) to satisfy this criterion. Due to the exclusive nature of $E_{PBX}$, only 1,853 of the original 4,119 materials will satisfy this criterion.

 We adapted our third selection criterion based on the selection criterion from the WhereWulff\cite{Sanspeur2023} high-throughput screening workflow for oxide catalyst discovery. Here, we assess the surface energy of every termination for each facet of each material that is Pourbaix stable with Equation~\ref{eq:surface_energy_chempot}. From the surface energy, we were able to construct the Wulff shape which indicates the most prominent facets in an equilibrium crystal. When considering which surfaces are stable and can potentially facilitate \gls{OER}, we only consider surfaces that appear on the Wulff shape. Depending on the stoichiometry of the slab, the surface energy can vary as a function of $\Delta \mu_M$. For simplicity, we will roughly assume a possible chemical potential range of the metal M as $-5 < \Delta \mu_M < 0$ eV when interpreting surface energy. Wulff shapes containing negative surface energies are ignored as nonphysical solutions as this indicates the surface is more stable than the bulk (which implies dissolution of the solid). The dissolution of the Wulff shape will consequently lead to 314 additional materials being omitted from our list of candidates, leaving us with 1,539 materials with 11,918 stable surfaces on the Wulff shape.  

In our fourth criterion, we assess the overpotential of each candidate surface with Equation~\ref{eq:overpotential}. We consider any material with at least two facets on the Wulff shape exhibiting $\eta < 0.75$ V as being potentially competitive with \ce{RuO2} and \ce{IrO2} in regards to catalytic performance. We find that 101 materials (512 surfaces) from our previous 1,539 materials will satisfy this criterion.

The fifth criterion assesses the thermodynamic stability of the candidate material via the energy above hull ($E_{hull}$) or the formation energy of a material relative to the ground state. Like with Pourbaix stability, materials with a calculated $E_{hull} > 0$ eV per atom are metastable with the likelihood of experimental synthesizeability decreasing as $E_{hull}$ increases. Materials with a calculated $E_{hull} \leq 0.1$ eV per atom have been shown to have reasonable rates of demonstrated synthesis in experiment\cite{Aykol2018}. From our aforementioned 101 catalytically active candidates, we identify 92 materials that are metastable. \color{black}

The final criterion assesses the material cost of each compound in \$ per kg. To satisfy this criteria, the cost of a compound must be less than that of \ce{RuO2} or \$18,315 per kg (as of March 2021\cite{dailymetalprice, apmex, metalary}). As a conservative estimate, we will also assume a price variation of \$9,969/kg (based on the lowest and highest prices of \ce{Ru} in the last 24 years) in the future and as such, all materials must be less than \$8,346/kg. We have identified 81 materials and 66 distinct chemical systems from our 101 metastable materials that have satisfied this criterion. A tabulated list of all candidates with the lowest value of $\eta$, $E_{hull}$, $E_{PBX}$, material cost, space group, and number of facets with low overpotentials is give in Table S1-S8\color{black}.  

\subsection{Overpotential assessment}

\begin{figure}[ht!]
    \centering
    \includegraphics[width=\linewidth]{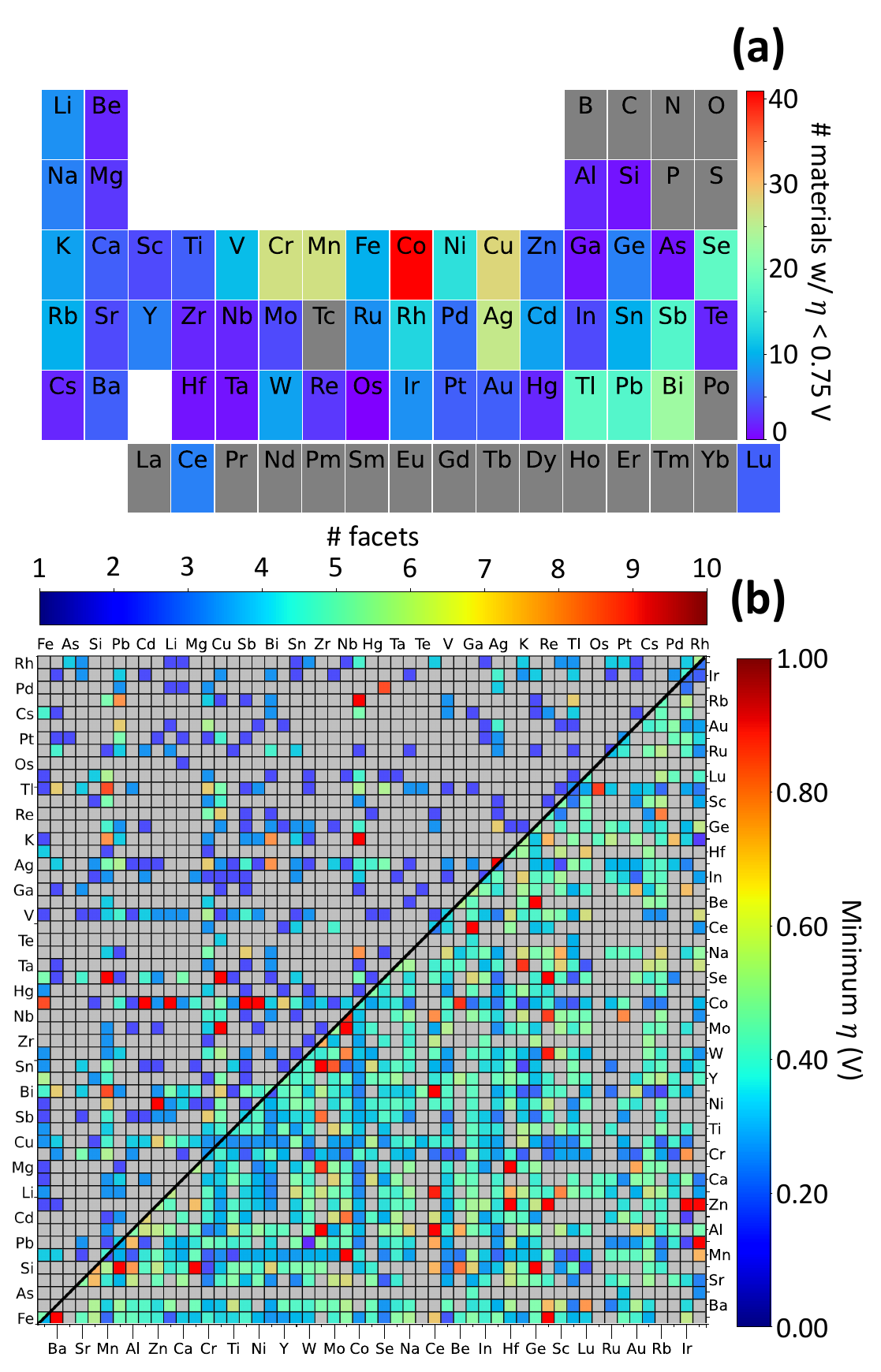}
    \caption{(a) Periodic table of elements considered in the database. Colormap indicates the number of materials containing the element that exhibit $\eta < 0.75$ V for at least two facets. (b) Grid map for each pair of elements with colors indicating the lowest overpotential amongst all facets on the Wulff shape containing this chemical system (bottom right) and the number of facets on the Wulff shape across material with the same chemical system that exhibit overpotentials less than 0.75 eV (top left). Tick labels on the x- and y-axis are sorted from the cheapest (Fe) to the most expensive (Rh) element.\color{black}}
    \label{fig:grid_overpotentials}
\end{figure}

Although our set of tiered screening criteria will provide the most economically and industrially viable candidates with respect to thermodynamic stability and material cost, we posit that there is a wider range of materials that can have competitive overpotentials necessary for \gls{OER} when ignoring the criteria listed above. Figure~\ref{fig:grid_overpotentials}(a) shows the 53\color{black}~elements on the periodic table considered in our database. The heatmap indicates the number of materials containing an element that exhibits at least 2 facets with $\eta < 0.75$ V. Oxide combinations containing Cr, Mn, Co, Cu, Bi, Pb, Se, Tl, Ag, and Sb\color{black}~tend to form catalysts that exhibit low overpotentials. Ag-based chemical systems such as Ag-O\cite{Joya2016}, Ag-Bi-O\cite{Simondson2022}, Ag-Cu-O\cite{Huang2015}, and Ag-Co-O\cite{Yu2020} have been demonstrated as viable catalysts for \gls{OER} in experimental settings. The common use of Ag has been attributed to the fact that Ag has the highest electrical conductivity amongst all metals, allowing it to easily facilitate the four electron charge transfer process that takes place during \gls{OER}\cite{Yu2020}. Although it is not as common and cheap as the 3d transition metals, Ag is still more reasonably priced than the other noble metals. Furthermore, it has been shown that Ag-doping at 1\% has been enough to enhance charge transfer in \gls{OER}\cite{Zhao2017}, making Ag-based oxides reasonably economical. Similarly, Mn-based chemical systems are known to be promising catalysts for \gls{OER}. This is owed to the intrinsically high activity and number of polymorphs for compounds of \ce{MnO_x}\cite{Tian2020} which can be synergistically improved when introducing other components such as Fe\cite{Teng2018}, Ni\cite{Hong2020}, and Co\cite{Hu2018}. Antimonates (Sb-based oxides) have been extensively studied in both computational\cite{Gunasooriya2020} and experimental\cite{Zhou2022a, Zhou2022b, Simondson2022, Luke2021} settings as promising low-cost anode catalysts for \gls{OER} owed to their low overpotential and high operational stability as a consequence of Sb-O p-d hybridisation\cite{Luke2021}. Mixed metal Co oxides are abundant in \ce{CO^{2+}} cations that are useful for \gls{OER}. The large atomic difference between Co and the larger metal in Co-based Perovskites can also lead to distortions in the structure that can better stabilize either the \ce{CO^{2+}} or \ce{CO^{3+}} cation which can allow for tunable active sites\cite{Gupta2023}.\color{black}

Non-noble chemical systems such as Fe-Ni\cite{Si2017}, Cu-Fe, Fe-Mn\cite{Li2015}, Fe\cite{Mullner2019}, Co-Mn\cite{Hirai2016}, and Co-Cr\cite{Al-Mamun2016}\color{black}~based oxides have been shown to exhibit competitive overpotentials in the experimental literature despite their absence from our final set of candidates. To account for this discrepancy, we predicted the overpotentials for all Wulff shape facets of all 4,119 materials. We summarized the overpotential of each chemical system by plotting the lowest overpotential and the number of facets exhibiting $\eta < 0.75$ eV in Figure~\ref{fig:grid_overpotentials}(b) across all materials of the same chemical system as a heat map. Upon further inspection, Fe-Mn-O\color{black}~is shown to be relatively unstable in our dataset ($E_{Pbx} = 0.95$ eV)\color{black}~despite having a competitive overpotential corroborating with past experiments. However, this was explored in the context of nanoparticle catalysts\cite{Li2015}\color{black}. %Our predictions also demonstrate $\eta < 0.75$ eV for Mn-O, Co-Fe-O, Co-Zn-O, and Co-Cr-O on facets across different materials of the same chemical system. These low overpotentials across surfaces of different materials can imply improved catalytic activity when assessing multicrystalline or even amorphous systems as has been observed for Co-Fe-O\cite{Indra2014}. 

% Figure~\ref{fig:periodic_table_materials} shows the 36 elements on the periodic table considered in our database. The heatmap indicates the number of materials containing an element that exhibits at least 2 facets with $\eta < 0.75$ V. Oxide combinations containing alkali metals, especially those containing \ce{K}, tend to exhibit low overpotentials when metal concentration is allowed to vary. Typically, these components do not make up the bulk composition of the catalyst (with the exception of perovskite catalysts), however the concentration of alkali cations in solution have been shown to improve \gls{OER} activity in some studies\cite{Jamesh2018, DelRosario2021, Yang2020}. \citet{DelRosario2021} demonstrated that the presence of various alkali in solution can improve the activity of \ce{IrO_x} and \ce{NiCo2O3} by 4- and 8-times respectively with \ce{K} exhibiting the greatest effect on improved activity. This is a result of the formation of \ce{M^+}-\ce{OH} complex that keep \ce{OH} from dissociating while transferring electrons to \ce{OOH^*} to form the intermediate.

We have demonstrated 380\color{black}~chemical systems exhibiting facets on the Wulff shape that are potentially active for \gls{OER} despite only 76\color{black}~chemical systems (92 materials) (ignoring material cost) appearing in our set of candidates due to the\color{black}~bulk stability of many materials being inaccessible. Methods for stabilizing bulk oxides such as nanoscale stability (e.g. in the aforementioned case for Fe-Mn-O\cite{Li2015})\color{black}, elemental doping, and the introduction of oxygen vacancies have been demonstrated to be effective means of improving stability\cite{Jamesh2018}. \ce{RuO2} for example is known to have stability issues in aqueous environments despite being the benchmark for \gls{OER} catalysts. However doping with other metals such as \ce{Cr} and \ce{Ti} has been shown to stabilize this material\cite{abed2023multi}. With these methods of stabilizing the catalyst material and accessing active surfaces, we emphasize that the identification of viable catalysts should not strictly be confined by the Pourbaix stability or even $E_{hull}$. As an example, we will demonstrate how nanoscale stabilization can potentially expand the material space available for \gls{OER} to access these materials.
\subsection{Alternative screening frameworks}

\begin{table}[ht!]
\caption{\label{tab:screening_framework_candidates} The different screening frameworks assessed in this study with varying reaction conditions and criteria for Pourbaix stability. All screening frameworks are assumed to occur under pH=1. Superscript letters are used to label each framework evaluated (see Table S1-S8 for a list of candidates that were identified in each framework). Numbers in parentheses correspond to conductive candidates with small band gaps ($E_{gap} < 0.1$ eV) Framework \textit{j} and \textit{k} correspond to the frameworks investigated in Figure~\ref{fig:screening}}
\scalebox{0.8}{
\begin{tabular}{c|c|c|c|c}
% \begin{tabular}{p{3cm}p{1cm}p{1cm}p{1cm}p{1cm}}
\hline
\hline
Temperature ($^oC$) & 60 & 60 & 80 & 80 \\
Applied Potential (V) & 1.8 & 1.2-2.0 & 1.8 & 1.2-2.0 \\
\hline
\hline
Bulk            & 122$^a$ & 99$^e$  & 120$^i$ & 99$^m$ \\
Bulk/Wulff      & 83$^b$  & 62$^f$  & 81$^j$  & 62$^n$ \\
Bulk/Wulff/Nano & 111$^c$  & 83$^g$  & 121$^k$ & 84$^o$ \\
Bulk/Nano       & 168$^d$ & 129$^h$ & 181$^l$ & 129$^p$ \\
\hline
\hline
\end{tabular}
}
\end{table}

Table~\ref{tab:screening_framework_candidates} shows the number of final candidates identified under different reaction conditions and definitions of Pourbaix stability. In this study we constructed a database of machine learning data and demonstrated how it can be used in a variety of ways to screen for catalyst. First, we identified potential candidates for \gls{OER} by creating a screening framework based on simple bulk thermodynamic (the energy above hull and Pourbaix energy above hull) and surface stability arguments  at $pH=1$, $U=1.8$ V, and $T=$ 80$^o$C (see \textit{j} in Table~\ref{tab:screening_framework_candidates}). We can easily modify our framework to reflect other reaction conditions and criteria for Pourbaix stability as shown in Table~\ref{tab:screening_framework_candidates}. For example, we can \color{black}~modify our framework by introducing an additional layer of complexity to our definition of Pourbaix stability by analyzing the possibility of nanoscale stability under acidic conditions for \gls{OER} as demonstrated in Figure~\ref{fig:screening} (right). Using our nanoscale stability diagrams (see Figure~\ref{fig:nanoscale_stability}), we were able to identify 2,778 Pourbaix stable materials with 886 stabilizing at the nanoscale regime (10 to 100 nm). From our Wulff shape analysis of the 2,778 Pourbaix stable materials, we identified 18,501\color{black}~surfaces that appear on the Wulff of 2,430\color{black}~materials. We find that 816\color{black}~of these surfaces (from 159\color{black}~materials) also exhibit low overpotentials. From these 159\color{black}~potentially active materials, 147\color{black}~are metastable and 121\color{black}~exhibit a material cost less than \ce{RuO2}. In total we have identified 121 candidates and 40 additional candidates that are potentially commercially viable for \gls{OER} when \color{black}~synthesized as nanoparticles (see \textit{k} in Table~\ref{tab:screening_framework_candidates}). 

Another example of how we can modify our screening framework is by changing our criteria for surface stability. As mentioned before, we only considered facets that appear on the Wulff shape within a metal chemical potential between $-5 < \Delta \mu_M < 0$ eV. The Wulff shape indicates the most statistically likely facets to appear under equilibrium crystal growth conditions. However, non-equilibrium conditions can potentially force different types of facets to appear\cite{Quan2013}. We can account for this in our screening framework by considering all facets as viable surfaces for \gls{OER}. With this simple assumption, we identified 181 materials (120 and 61 \color{black}~in the bulk and nanoscale regime respectively) that satisfy all criteria (see \textit{l} in Table~\ref{tab:screening_framework_candidates}). Among the chemical systems identified as nanostable is Fe-Mn-O which as mentioned previously, has been explored as nanoparticle catalysts for \gls{OER}\cite{Li2015}.\color{black}

Other alternative screening frameworks can be explored by simply adding or modifying existing criteria. As a simple example, we can assess Pourbaix stability under different temperature conditions. The operating temperatures for \gls{OER} are typically between 60 $^o$C and 80 $^o$C with higher temperatures resulting in improved ionic conductivity and kinetics in exchange for lower stability. Although our assessment assumes an operating temperature of 80 $^o$C, we can easily re-assess the second criteria (for bulk stability) in our screening framework at 60 $^o$C instead which will yield 2 additional candidates without considering nanoscale stability (\textit{a} and \textit{b}). When nanoscale stability is considered however, lowering the temperature from 80 $^o$C to 60 $^o$C will result in a loss of 10 (\textit{c}) and 13 (\textit{d}) candidates with and without Wulff shape stability respectively. The bulk Pourbaix formation energy of ground state polymorphs decreases at lower temperatures. This decreases the likelihood for metastable polymorphs to become more stable than the ground state at the nanoscale resulting in fewer nanostable candidates at lower temperatures.\color{black}

We can also consider more complex and realistic reaction conditions whereby candidates are assessed based on their stability over a range of operating potentials. We adopted the potential range of 1.2-2.0 V for \gls{OER} for which a material must remain stable (or metastable with $E_{PBX} \leq 0.5$ eV) as suggested by~\citet{Wang2020} (\textit{e-h} at $T=$ 60 $^o$C and \textit{m-p} at $T=$ 80 $^o$C). This strict set of operating conditions unsurprisingly results in fewer candidates when compared to a static potential operating condition of 1.8 V. The total number of candidates presented in this work however, still far exceeds the original set of candidates identified by~\citet{Wang2020} and subsequently proposed by~\citet{Gunasooriya2020}. The \gls{OC22} framework is trained on data calculated with the \gls{PBE}-\gls{GGA} functional whereas the data presented by~\citet{Wang2020} was obtained using the more accurate SCAN functional\cite{Wang2020b} with the addition of correction terms better suited for assessing stability under corrosive conditions. These considerations contributed to a more realistic, albeit pessimistic, set of 17 Pourbaix stable and active candidates subsequently found by~\citet{Gunasooriya2020} as oppose to the 99 bulk stable candidates identified at 80 $^o$C (\textit{m}). In a future work, we hope to adapt these functionals and corrections when assessing nanostable catalysts to provide a more accurate expansion to the limited set of bulk stable candidates explored by the references herein~\cite{Wang2020, Gunasooriya2020}\color{black}.

In total we have identified 190 candidates (122 bulk- and 68 nanostable) with 145 distinct chemical systems when considering all the different screening frameworks listed in Table~\ref{tab:screening_framework_candidates}. In the next section, we will validate our findings by comparing to past experimental and computational results as well as our own \gls{DFT} simulations.\color{black}~We hope this demonstration regarding the ease and variability of how these frameworks can be modified illustrates the utility of these machine learning databases. We highly encourage the scientific community to use our database to explore further possibilities and alternative screening frameworks in the future. 
\subsection{DFT and literature validation}

\begin{figure*}[ht!]
    \centering
    \includegraphics[width=1\linewidth]{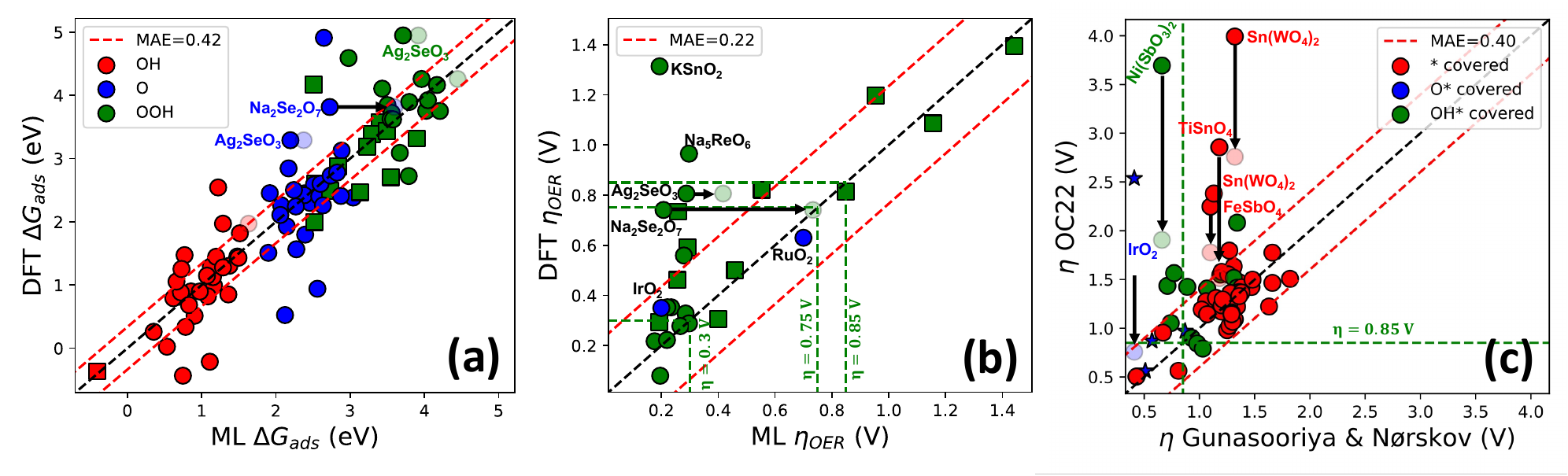}
    \caption{(a) \gls{DFT} calculated data points for reaction energy plotted against the corresponding \gls{ML} predicted quantities. (b) \gls{DFT} calculated data points for overpotential plotted against the corresponding \gls{ML} predicted quantities. Overpotentials of the benchmark materials are highlighted in blue (\ce{IrO2} and \ce{RuO2}). Overpotentials of less than 0.75 eV are considered catalytically active. Square data points indicate desorption or dissociation of the intermediate. Transparent data points correspond to \gls{ML} predicted data points with spring constraints applied to prevent desorption and dissociation events.}
    \label{fig:validation}
\end{figure*}

We used \gls{DFT} to validate 85 values for Gibbs adsorption energy corresponding to Equations~\ref{eq:gads1_corr}-\ref{eq:gads3_corr} for 33 Pourbaix stable compounds in our database. The \gls{MAE} of the test set shown in Figure~\ref{fig:validation}(a) is 0.42 which is larger than the \gls{MAE} of 0.239 eV obtained from the validation set in the original \gls{OC22} assessment\cite{Tran2022b}. We find no difference in the amount of error when validating \gls{ML} data points with (square) and without (circle) desorption/dissociation event. However, \gls{ML} values corresponding to \ce{O^*} adsorption on \ce{Ag2SeO3} and \ce{Na2Se2O7} and \ce{OOH^*} adsorption on \ce{Ag2SeO3} will underestimate $\Delta G_{ads}$ relative to \gls{DFT}. These data points exhibit better agreement with \gls{DFT} when applying a spring constant constraint to prevent dissociation/desorption events (transparent data points). This implies that the model from \gls{OC22} may overestimate the tendency for desorption/dissociation in some combinations of intermediates and surface and could be a potential point of improvement in future iterations of the model.

Next we assess the predictability of the \gls{ML} inferred overpotential (Equation~\ref{eq:overpotential}) using the predicted and calculated values of $\Delta G_{ads}$ from Figure~\ref{fig:validation}(a). Figure~\ref{fig:validation}(b) once again plots the \gls{DFT} calculated data points against the corresponding \gls{ML} quantities. The majority of data points sampled will lie within the \gls{MAE} of 0.22 V, however overpotentials corresponding to \ce{KSnO2}, \ce{Na5ReO6}, \ce{Ag2SeO3}, and \ce{Na2Se2O7} will lie outside the \gls{MAE}. \ce{Ag2SeO3} and \ce{Na2Se2O7} will exhibit overpotentials closer to parity with \gls{DFT} when applying a spring constant constraint to prevent the dissociation/desorption of \ce{O^*} and \ce{OOH^*} in \ce{Ag2SeO3} and \ce{O^*} in \ce{Na2Se2O7} as shown in Figure~\ref{fig:validation}(a). Despite the large deviation from \gls{MAE} however, we find that most data points with predicted low overpotentials will still exhibit low overpotentials close to or less than 0.75 V using \gls{DFT} with the exception of \ce{KSnO2} and \ce{Na5ReO6}. 

From our \gls{DFT} calculations, we were able to identify 6 data points with overpotentials below the soft theoretical limit ($\eta_{Th} > 0.3$ V) imposed by scaling relationships: \ce{Mn23FeO32}, \ce{HgSeO4}, \ce{Na2Co2O3}, \ce{Cd2PbO4}, \ce{MnTlO3}, and \ce{KBiO2}. The Mn-Fe-O chemical system is well explored in the literature with experimental overpotentials as low as 0.47 V\cite{Li2015, Konno2021, Han2022}. Although no studies have investigated ordered structures of Na-Co-O, the doping of layered \ce{CoO2} with Na has resulted in overpotentials as low as 0.24 V\cite{Sun2021}. Although we have predicted \ce{HgSeO4} and \ce{Cd2PbO4} as having overpotentials less than $\eta_{Th}$, Pb, Cd, and Hg are known to possess potential health risks\cite{osha} and as such, caution is advised for any further investigation of candidates containing these elements. As far as we are aware, the Mn-Tl-O and Bi-K-O chemical systems have yet to be explored in the literature.

Lastly, we compared our predicted values of overpotential to those obtained by~\citet{Gunasooriya2020} for various facets of 11 materials in Figure~\ref{fig:validation}(b). The majority of our predicted datapoints lie within a \gls{MAE} of 0.38 V. \gls{ML} data points tend to underestimate the \gls{DFT} values. However, those that lie well beyond 0.38 V, such as \ce{IrO2}, \ce{Ni(SbO3)2}, \ce{TiSnO4}, \ce{Sn(WO4)2}, \ce{Sn(WO4)2}, \ce{FeSbO4} will again exhibit overpotentials much closer to parity with \gls{DFT} when applying a spring constant constraint to prevent the dissociation/desorption. A key difference in the methods used to obtained both datasets is that the surfaces observed by~\citet{Gunasooriya2020} can be covered completely by \ce{O^*} or \ce{OH^*} whereas all surfaces considered in this study only considered the adsorption of a single \ce{O^*} or \ce{OH^*} intermediate at a time. Despite this, we still find many data points terminated by \ce{O^*} or \ce{OH^*} within the \gls{MAE}.

Table S9-S14 lists all 190 candidate materials we have identified in this work along with references to the experimental literature where available and the \gls{PDS} of the surface with the lowest overpotential. In total, we have identified 102 out of 145 unique chemical systems (127 out of 190 materials) that have yet to be explored with 77 non-toxic (does not contain Pb, Cd, Hg, or Cr) chemical systems (98 materials). Although not considered in our study, conductivity is also required for the operation of electrocatalysts\cite{Wang2020}. Of these 98 non-toxic and unexplored candidates, 27 have a band gap of less than 0.1 eV according to the Materials Project\cite{Jain2013} which satisfies this additional criteria. These conductive candidates are: \ce{Cu3(SbO3)4}, \ce{Ba8(Bi2O7)3}, \ce{BaBiO3}, \ce{Cu3Mo2O9}, \ce{Mg(BiO3)2}, \ce{CuMoO4}, \ce{Li(Bi3O5)4}, \ce{Ce2Mo4O15}, \ce{Ce9YO20}, \ce{AgSnO3}, \ce{Ce2(GeO3)3}, \ce{Ag4GeO4}, \ce{MnTlO3}, \ce{LuCoO3}, \ce{BaMn2O3}, \ce{Li(CuO)2}, \ce{MnBiO3}, \ce{Mn2BeO4}, \ce{VZn2O4}, \ce{ScMn2O4}, \ce{LiMn3O4}, \ce{Mn3NiO4}, \ce{Mn2NiO3}, \ce{VSbO4}, \ce{Ag3RuO4}, \ce{TiCu3O4}, and \ce{TlCuO2}. The other 71 candidates, although non-conducting, can potentially be considered as anodes in photocatalytic mechanisms for \gls{OER}. The discovery of a potential candidate for \gls{OER} demonstrates the potential of \gls{ML}-assisted screening techniques in identifying novel catalysts.

\color{black}
\section{Conclusions}
In this contribution, we employed pre-trained machine learning potentials from the Open Catalyst Project to develop a publicly available database of \ce{O^*} and \ce{OH^*} surface intermediates for 4,119 oxide materials. %This database required 9,473.9 GPU hours to complete and contains 5,969,157 total \gls{DFT} energy predictions with 183,560 bare slab and 5,785,597 surface intermediate predictions for all terminations of facets with a maximum Miller index of 1. 
We demonstrated the utility and variability of this database by presenting several easily implemented high-throughput screening frameworks for identifying thermodynamically stable and catalytically active (low overpotential) materials for oxygen evolution reaction from our initial pool of 4,119 candidates. Our first screening framework identified 81\color{black}~candidates that are Pourbaix stable in the bulk regime with catalytic surfaces that appear on the Wulff shape. By slightly modifying this framework to account for the possibility of nanoscale stability, we identified 121 additional candidates. Additional modifications to the reaction conditions and our definition of Pourbaix stability yields a total of 190 candidates with 27 unexplored candidates fulfilling additional criteria of being non-toxic and conductive. Furthermore, we were able to validate our predictions with \gls{DFT} calculations from the literature as well as our own. When ignoring material cost and bulk stability, we find that oxides containing Cr, Mn, Co, Cu, Se, Sb, Bi, Pb, and Tl\color{black}~tend to exhibit overpotentials that are possibly competitive with current benchmark materials (\ce{IrO2} and \ce{RuO2}). We plan to further expand our database to include other potential reaction mechanisms for OER such as the oxo-coupling mechanism and lattice oxygen evolution. We hope this database will encourage future investigators to develop their own high-throughput screening frameworks.

\section*{Author Contributions}
Richard Tran implemented the screening framework and performed DFT calculations. Liqiang Huang designed the parameters and software to manage and automate predictions and performed ML predictions. Yuan Zi, Shengguang Wang, and Benjamin M. Comer defined the overall goal of the project and provided scientific insight and solutions. Lars Grabow defined the DFT aspect of the work. Jiefu Chen and Xuqing Wu defined machine learning aspect of the work. Jiefu Chen and Lars Grabow acquired funding for the project. Ligang Lu initiated and funded the project and recruited Shell Subject Matter Experts to the project. Ligang Lu, Stefan J. Raaijman, Nishant K. Sinha, Sajanikumari Sadasivan, Shibin Thundiyil, Kuldeep B. Mamtani, and Ganesh Iyer defined the industrial benchmarks and goals of the project. All authors contributed to and commented on the manuscript.

\section*{Conflicts of interest}
This project is supported by Shell US.

\section*{Data Availability}
Data for this article, including including the \gls{ML} predicted initial and relaxed structures and total energies are available at the University of Houston Dataverse Repository at \href{https://dataverse.tdl.org/dataset.xhtml?persistentId=doi:10.18738/T8/APJFTM}{https://doi.org/10.18738/T8/APJFTM}\cite{APJFTM_2023}. Details regarding the database metadata are given in the Supplementary Information. The code for all analysis and data generation can be found on \href{https://github.com/moxx799/Shell_repo}{Github} with DOI: 10.5281/zenodo.12616291.

\section*{Acknowledgements}
The authors gratefully acknowledge financial support from Shell US. The authors also gratefully acknowledge financial support from the University of Houston Energy Transition Institute. This research was performed using computational resources from the Carya, Opuntia, and Sabine clusters provided by the Research Computing Data Core at the University of Houston.
\bibliography{rsc}

\clearpage
\onecolumn
\section{Supplementary Information}
\subsection{Surface energy}

To the surface energy of all bare slabs, we begin with the surface grand potential given by:
\begin{equation}
\label{eq:surface_grand_potential}
\gamma = \frac{E_{slab} - \sum_i n_i\mu_i}{2A}
\end{equation}
where $E_{slab}$ is the total energy of the bare slab, $n_i$ is the number of atom $i$ in the slab, $\mu_i$ is the chemical potential of atom $i$, $A$ is the surface area and the factor of $\frac{1}{2}$ accounts for the two surfaces in the slab. We can define the chemical potential of each element as:
\begin{equation}
\label{eq:bulk_chempot}
\mu_{\ce{A_xB_yO_z}} = E^{\ce{A_xB_yO_z}}_{bulk} = x\mu_A + y\mu_B + z\mu_O
\end{equation}
where $E^{\ce{A_xB_yO_z}}_{bulk}$ is the total energy per formula unit of the bulk crystal. As such, the surface energy for stoichiometric slabs can be rewritten as:
\begin{equation}
\label{eq:surface_energy}
\gamma = \frac{E^{\ce{A_{nx}B_{ny}O_{nz}}}_{slab} - nE^{\ce{A_xB_yO_z}}_{bulk}}{2A}
\end{equation}
where n is the number of bulk formula units in the slab. 

We reiterate that we enumerated through all terminations per facet by modelling slabs with symmetrically equivalent terminations on each side. Inevitably, this will require the removal or addition of cations and oxygen which will lead to non-stoichiometric (relative to the bulk) slab models. In such cases, we need to compensate for the excess or deficient components by introducing variable chemical potentials per component. For any slab, there can be up to $n-1$ excess or deficient components relative to the bulk. As an example, the surface energy of a slab of \ce{A_{nx}B_{ny+k}O_{nz-j}} constructed from a bulk crystal of \ce{A_{x}B_{y}O_{z}} becomes:
\begin{equation}
\label{eq:surface_energy_chempot_SI}
\gamma = \frac{E_{slab}^{\ce{A_{nx}B_{ny+k}O_{nz-j}}} - (nE^{\ce{A_xB_yO_z}}_{bulk} + k\mu_B - j\frac{1}{2}\mu_{O_2})}{2A}
\end{equation}

The chemical potential of oxygen ($\mu_O$) can be referenced to the electrochemical decomposition of water to $\ce{O2_{(g)}}$:
\begin{equation}
\label{eq:water_breakdown_SI}
2\ce{H2O_{(g)}} \rightarrow 4(\ce{H^+} + e^-) + O_{2(g)}:\qquad 4.92~eV
\end{equation}
given by:
\begin{equation}
\label{eq:oxygen_chempot}
\mu_{O_2} = 4.92 + 2G_{\ce{H2O}} - 4(\mu_{\ce{H^+}} + \mu_{e^-}) + \Delta G^{O^*}_{corr}
\end{equation}
where
\begin{equation}
\label{eq:g}
G = E + ZPE - TS^o
\end{equation}
We can relate the proton-electron pair ($H^+ + e^-$) to the activity of the proton and the \gls{SHE} using the Nernst Equation:
\begin{equation}
\label{eq:nernst_eqn}
\mu_{H^+} + \mu_{e^-} = \frac{1}{2}G_{\ce{H2}} - eU +k_BTlna_{H^+}
\end{equation}
Here $a_{H^+}$ is the activity of a proton with $-pHln10=lna_{H^+}$ and $eU$ is the change in electron energy under an applied potential. An excess or deficient oxygen component in the slab can be treated as an adsorbed or desorbed species and to account for this, we included a correction term, $\Delta G^{O^*}_{corr}$, to derive the Gibbs free energy of adsorption from the \gls{DFT} electronic \ce{O^*} adsorption energy (see~\citet{Tran2022b} and~\citet{Gunasooriya2020} for details in regards to corrections made for the Gibbs free energy). Consequently, $\mu_O$ can be rewritten as a function of $pH$ and $U$ as such:
\begin{equation}
\label{eq:muO_pH_U_SI}
\mu_{O_2} = 4.92 + 2\mu_{\ce{H2O}} - 4(\frac{1}{2}\mu_{\ce{H2}} - eU +k_BTlna_{H^+}) + \Delta G^{O^*}_{corr}
\end{equation}

It is typical to reference the chemical potential of B ($\mu_{B}$) with respect to the per atom energy of the ground state bulk crystal of pure component B (e.g. $\mu_{Fe} = \Delta \mu_{Fe}+E^{DFT}_{BCC, Fe}$). However, this is predicated upon the assumption that \ce{A_xFe_yO_z} will immediately decompose into pure solid BCC Fe at the surface. However, multicomponent systems generally do not immediately decompose into individual solid components of each element. Herein, we assume that an excess or deficiency of components \ce{B} or \ce{O} at the surface will lead to slight surface passivation with a decomposition of \ce{A_xB_yO_z} into a more stable multicomponent guided by the phase diagram, e.g.:
\begin{equation}
\label{eq:example_of_decomp}
\ce{A_xB_yO_z} \rightarrow xA + \ce{B_yO_z}
\end{equation}
In such a case, it makes sense to reference $\mu_B$ with respect to the energy of \ce{B_yO_z}:
\begin{equation}
\label{eq:chempot_of_B}
\mu_{\ce{B_yO_z}} = E^{\ce{B_yO_z}}_{bulk} = y\mu_B + z\mu_O
\end{equation}
Here, the chemical potential of B increases (or decreases) with the component of B at the surface, leading to a slight passivation of \ce{A_xB_yO_z} to \ce{B_yO_z} at the surface. We can thereby rewrite $\mu_B$ as a function of $\mu_O$
\begin{equation}
\label{eq:chempot_of_B_as_O}
\mu_B = \frac{E^{\ce{B_yO_z}}_{bulk} - z\mu_O}{y}
\end{equation}
which can be substituted into Equation~\ref{eq:surface_energy_chempot_SI} in order to define surface energy purely as a function of $\mu_O$ and by extension $U$ and $pH$ in accordance with Equation~\ref{eq:muO_pH_U_SI}. 

\subsection{Nanoparticle formation energy}

The nanoparticle formation energy ($G^{NP}_{f}$) is given by Equation 4 in the main manuscript. For a material in the bulk regime ($r > 100$ nm), we can assume that the thermodynamic contributions of the surface are negligible when compared to the contribution from the bulk. However, as the particle size continues to decrease, the surface-area-to-volume ratio will increase, resulting the properties of the surface dictating the properties of the overall material. Unlike the bulk formation energy ($E_V(pH, V, T)(\frac{4}{3}\pi r^3)$), $G^{NP}_{f}$ accounts for this by incorporating the surface energy contributions into the overall formation energy. 

Here, $\frac{4}{3}\pi r^3$ is the volume of a nanoparticle at radius $r$ and $E_V$ is the Pourbaix formation energy at a given pH, U, and T per unit cell volume ($E_{PBX}/V$)  where $E_{PBX}$ can be derived from the following\cite{Persson2012, Singh2017, Patel2019}:
\begin{equation}
\label{eq:bulk_pourbaix_formation_energy}
E_{PBX} = E_0 + k_BTln(10)pH - n_O \mu^o_{\ce{H2O}} + (n_H - 2n_O) pH + n_{e^-} (-n_H + 2n_O + eU)
\end{equation}
where $n$ are the number of species in the system respectively and $E_0$ is the formation energy of the bulk with respect to \ce{H2_{(g)}} and \ce{O2_{(g)}}:
\begin{equation}
\label{eq:bulk_formation_energy_wrt_gas}
E_0 = E_f + k_BTln(10)pH - n_{\ce{H2O}}\mu^o_{\ce{H2O}}
\end{equation}

The contribution of the surface energy is given by $\bar{\gamma}(4\pi r^2)$ where $4\pi r^2$ is the surface area of the nanoparticle at radius $r$ and $\bar{\gamma}$ is the surface energy of the nanoparticle. In this study, the equilibrium crystal structure, or Wulff shape, serves as an analogue to the nanoparticle. The Wulff shape is derived through the Wulff construction whereby a set of Miller index (hkl) planes perpendicular to a vector from an origin at a distance proportional to $\gamma_{hkl}$ (see Equation 1 in the main manuscript) enclose a polyhedron. The surface energy of this polyhedron is defined by:
\begin{equation}
\label{eq:weighted_surface_energy}
\bar{\gamma}(pH, U, T, \Delta\mu_M) = \frac{\sum_{hkl}\gamma_{hkl}(pH, U, T, \Delta\mu_M)A_{hkl}}{\sum_{hkl}A_{hkl}} = \sum_{hkl}\gamma_{hkl}f^A_{hkl}(pH, U, T, \Delta\mu_M)
\end{equation}
where $(hkl)$ are the facets that appear on the Wulff shape and $f^A_{hkl}$ is the fraction of area occupied by facet $(hkl)$ on the Wulff shape. 

Incorporating the surface energy contributions of the nanoparticle can consequently lead to some materials above the Pourbaix hull becoming more thermodynamically stable than the ground state material depending on the nanoparticle size.

\subsection{Overpotential}\label{overpotential_methods}

The overpotential has been demonstrated to be an excellent predictor of catalytic activity in electrocatalytic processes. The overpotential ($\eta$) describes the excess amount of applied potential required to move forward in each reaction step shown in Figure 3 relative to an ideal catalyst. The summation of each reaction energy (energetic height of each reaction step or $\Delta G_{rxn}$) is defined as the standard reduction potential in Equation~\ref{eq:water_breakdown_SI}. The ideal catalyst equally distributes the standard reduction potential along each step to minimize the required energy needed to move the reaction in the forward direction. This required energy is the equilibrium potential of \gls{OER} and is given as:
\begin{equation}
\label{eq:overall_energy}
\frac{G_{\ce{O2}} + 2G_{\ce{H2}} - 2G_{\ce{H2O}}}{4e}/ = 1.23 V
\end{equation}
As such, the theoretical overpotential for an electrocatalyst is given by:
\begin{equation}
\label{eq:overpotential_SI}
\eta^{OER} = max(\Delta G^{1}_{rxn}, \Delta G^{2}_{rxn}, \Delta G^{3}_{rxn}, \Delta G^{4}_{rxn})/e - 1.23 V
\end{equation}
where $max(\Delta G^{1}_{rxn}, \Delta G^{2}_{rxn}, \Delta G^{3}_{rxn}, \Delta G^{4}_{rxn})$ is the reaction energy of the potential determining step or $\Delta G^{RDS}_{rxn}$.

The energy of each reaction step is relative to the energy of a bare slab and two water molecules in a vacuum, all of which are initially assumed to be non-interacting. Guided by Equations i-iv (see Figure 3, we can determine the \gls{DFT} electronic adsorption energy of steps i-iii with the following:
\begin{equation}
\label{eq:eads1}
E^{\ce{OH^*}}_{ads} = E^{\ce{OH^*}} + \frac{1}{2}E_{\ce{H2}} - E_{slab} - \mu^o_{\ce{H2O}}
\end{equation}
\begin{equation}
\label{eq:eads2}
E^{\ce{O^*}}_{ads} = E^{\ce{O^*}} + E_{\ce{H2}} - E_{slab} - \mu^o_{\ce{H2O}}
\end{equation}
\begin{equation}
\label{eq:eads3}
E^{\ce{OOH^*}}_{ads} = E^{\ce{OOH^*}} + \frac{3}{2}E_{\ce{H2}} - E_{slab} - 2\mu^o_{\ce{H2O}}
\end{equation}
where $E^{X^*}$ is the total energy of the surface intermediate with adsorbate $X$ and $E_{X}$ is the reference energy of the adsorbate in a gas. By incorporating the vibrational frequency contributions of the adsorbate on the surface in $E^{ads*}$, i.e. the zero point energy and entropy, we can derive the corresponding Gibbs adsorption energies for steps i-iv:
\begin{equation}
\label{eq:gads1}
\Delta G^{i} = G^{\ce{OH^*}} + \mu_{H^+} + \mu_{e^-} - E_{slab} - G_{\ce{H2O}}
\end{equation}
\begin{equation}
\label{eq:gads2}
\Delta G^{ii} = G^{\ce{O^*}} + 2(\mu_{H^+} + \mu_{e^-}) - E_{slab} - G_{\ce{H2O}}
\end{equation}
\begin{equation}
\label{eq:gads3}
\Delta G^{iii} = G^{\ce{OOH^*}} + 3(\mu_{H^+} + \mu_{e^-}) - E_{slab} - 2G_{\ce{H2O}}
\end{equation}
\begin{equation}
\label{eq:gads4}
\Delta G^{iv} = 4.92 V = G_{\ce{O2}} + 4(\mu_{H^+} + \mu_{e^-}) - 2G_{\ce{H2O}}
\end{equation}

\begin{figure*}
    \centering
    \includegraphics[width=1\linewidth]{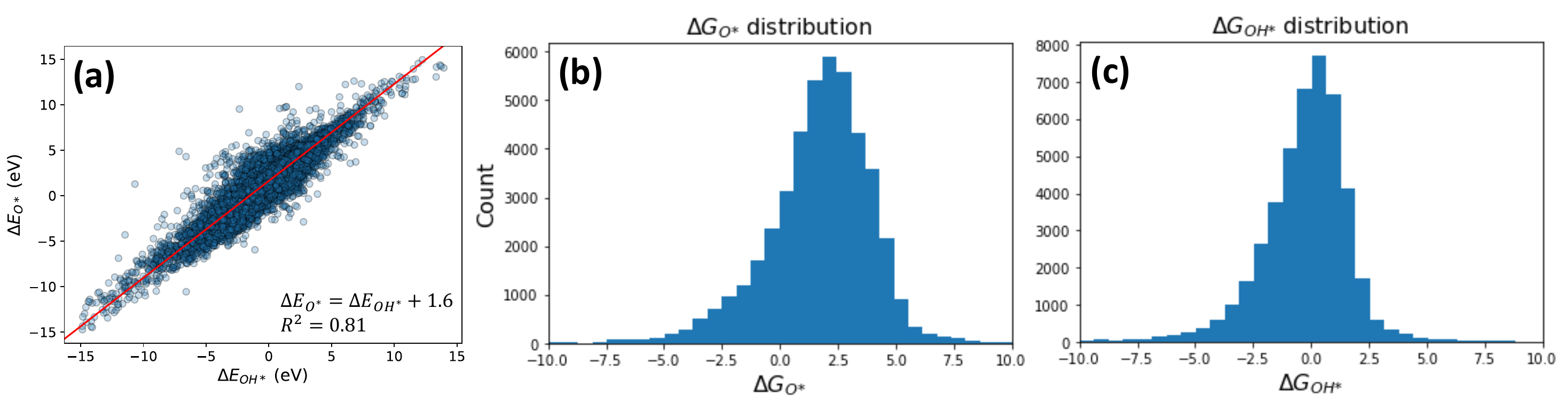}
    \caption{(a) Scaling relationship between $\Delta G_{OH^*}$ and $\Delta G_{O^*}$. Binding energies of $\Delta G > 15 eV$ or $\Delta G < -15 eV$ were omitted from scaling relationship as anomalies. Distribution of adsorption energies for $\Delta G_{O^*}$ (b) and $\Delta G_{OH^*}$ (c).}
    \label{fig:scaling_relationship}
\end{figure*}

As previously mentioned, a constant correction term ($\Delta G_{corr}$) can also be added to Equations~\ref{eq:eads1}-\ref{eq:eads3} to obtain $\Delta G$:
\begin{equation}
\label{eq:gads1_corr_SI}
\Delta G^{i} = E_{ads}^{\ce{OH^*}} + \Delta G^{OH^*}_{corr} + \mu_{H^+} + \mu_{e^-}
\end{equation}
\begin{equation}
\label{eq:gads2_corr_SI}
\Delta G^{ii} = E^{\ce{O^*}} + \Delta G^{OH^*}_{corr} + 2(\mu_{H^+} + \mu_{e^-})
\end{equation}
\begin{equation}
\label{eq:gads3_corr_SI}
\Delta G^{iii} = E^{\ce{OOH^*}} + \Delta G^{OH^*}_{corr} + 3(\mu_{H^+} + \mu_{e^-})
\end{equation}
where $G^{OH^*}_{corr}$, $G^{O^*}_{corr}$, and $G^{OOH^*}_{corr}$ were derived in the Supplementary Information of \gls{OC22}\cite{Tran2022b}. Alternatively, if $E^{OOH^*}$ is unavailable, well known scaling relationships between $\Delta G^{iii}$ and $\Delta G^{i}$ can be used instead (Figure S\ref{fig:scaling_relationship})):
\begin{equation}
\label{eq:gads3_scaled_SI}
\Delta G^{iii} = \Delta G^{i} + 3.26
\end{equation}

From Equations~\ref{eq:gads1}-\ref{eq:gads4}~we can then obtain the individual reaction energies list in Figure 1:
\begin{equation}
\label{eq:grxn1}
\Delta G^{1}_{rxn} = G^{i}
\end{equation}
\begin{equation}
\label{eq:grxn2}
\Delta G^{2}_{rxn} = G^{ii} - G^{i}
\end{equation}
\begin{equation}
\label{eq:grxn3}
\Delta G^{3}_{rxn} = G^{iii} - G^{ii}
\end{equation}
\begin{equation}
\label{eq:grxn4}
\Delta G^{4}_{rxn} = 4.92 - G^{iii}
\end{equation}

\subsection{Predictions of OOH* and dissociation and desorption events}

\begin{figure*}
    \centering
    \includegraphics[width=1\linewidth]{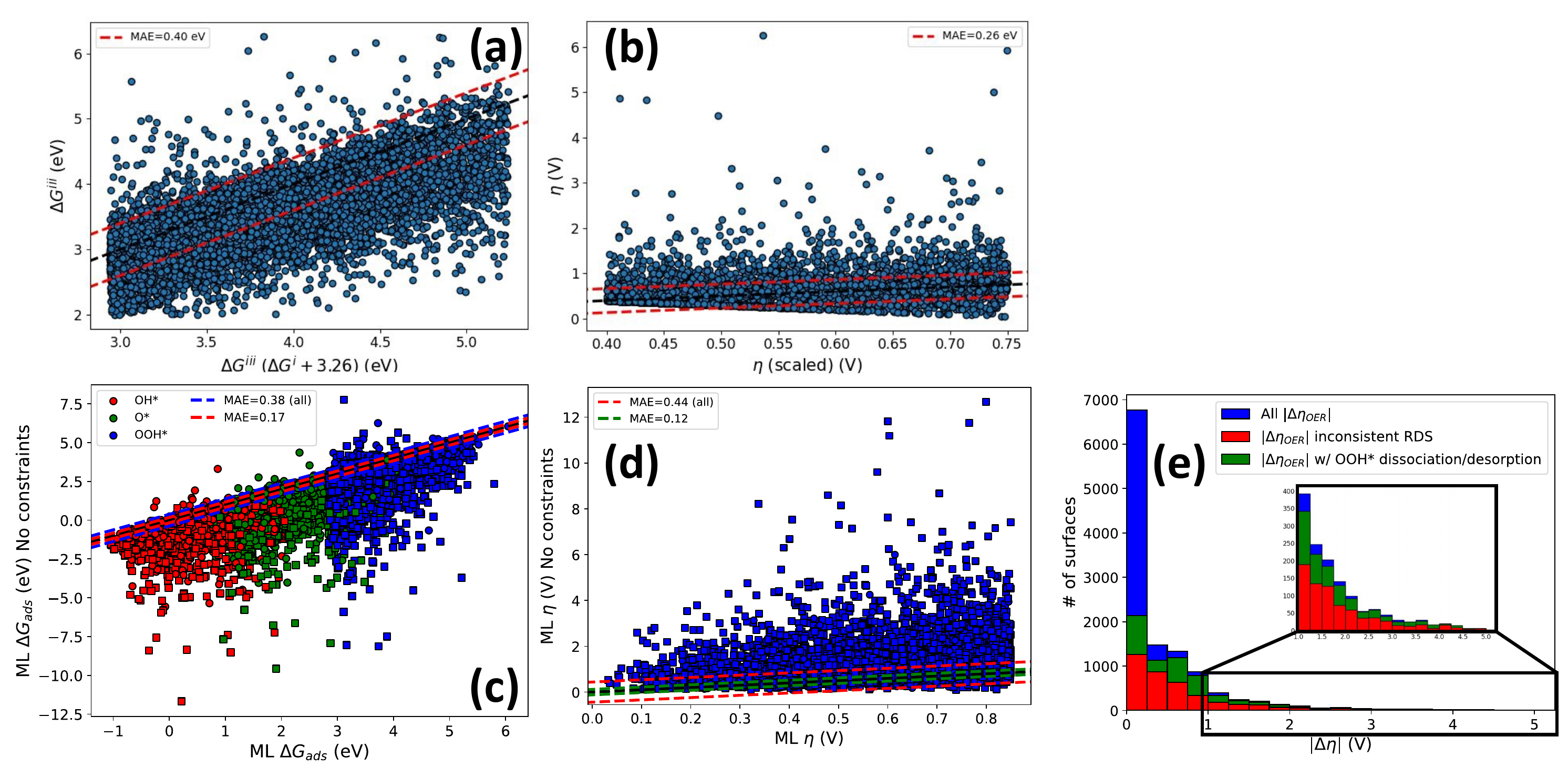}
    \caption{(a) Comparison of Gibbs free energy of adsorption for \ce{OOH^*} ($\Delta G^{iii}$) using Equation~\ref{eq:gads3_scaled_SI} (x-axis) vs $\Delta G^{iii}$ using Equation~\ref{eq:gads3} with \gls{ML} predictions of $E_{ads}^{\ce{OOH^*}}$ (b) and the corresponding data points for overpotential. (c) Comparison of the Gibbs free energy of adsorption ($\Delta G_{ads}$) with spring constraints to prevent adsorbate dissociation and desorption (x-axis) and without (y-axis) (d) and the corresponding data points for overpotential. Square data points indicate dissociation or desorption. The blue dashed line corresponds to an \gls{MAE} when all data points are considered and the red dashed line indicates an \gls{MAE} with dissociation and desorption events ommitted. (e) Distribution in the absolute difference in overpotential between values calculated with and without the aforementioned constraints with an inset for datapoints exhibit a difference greater than 1.0 V.\color{black}}
    \label{fig:ml_predictions_comparisons}
\end{figure*}

To save computational resources and time, we performed all initial assessments of $\eta$ for all surfaces by predicting the Gibbs free energy of \ce{OH^*} and \ce{O^*} with \gls{ML} and \ce{OOH^*} using the scaling relationship provided by Equation~\ref{eq:gads3_scaled_SI}. We acknowledge that this is a key intermediate in the \gls{WNA} mechanism and by avoiding this step, it is unclear if candidate catalysts will undergo alternative mechanisms for \gls{OER}. Furthermore, Equation~\ref{eq:gads3_scaled_SI} confines our exploration of overpotentials to the theoretical limit of 0.3-0.37 V regardless if any candidate studied can potentially break scaling relationships. Subsequently, we predicted the \ce{OOH^*} intermediate for all surfaces exhibiting $\eta < 0.75 V$. Figure S\ref{fig:ml_predictions_comparisons}(a) shows the \ce{OOH^*} Gibbs free energy ($\Delta G^{iii}$) of adsorption of these surfaces using Equation~\ref{eq:gads3_scaled_SI} (x-axis) and \gls{ML} predicted $E^{\ce{OOH^*}}_{ads}$. Although a large majority of data points lie within a \gls{MAE} of 0.4 eV (62\% or 5,861 out of 9,516 datapoints), we find that most data points beyond the \gls{MAE} illustrates the predicted $\Delta G^{iii}$ will severely underestimate the corresponding value obtained through a scaling relationship. We also find that the majority of overpotential data points (65\% or 2,324 out of 3,554 data points) assessed using both methods will consistently fall below 0.75 V, albeit with 2,368 within the \gls{MAE}.

In predicting the overpotential of each surface, we also initially applied a spring constant of $7.5 eV/Å^{-2}$ between all adsorbate atoms and $2 eV/Å^{-2}$ between the surface and adsorbate to avoid desorption and dissociation events. However, this could lead to final geometries far away from the energy minima, resulting in poor energy predictions and unreliable geometries. Subsequently, for all predictions of intermediates that contributed to a free energy diagram exhibiting $\eta < 0.75 V$, we performed additional \gls{ML} relaxation steps on the final predicted geometries (obtained with constraints induced by the spring constant) \textit{sans} the constraints. This two step relaxation process will help minimize the number of dissociation and desorption events when possible while providing the most reliable predicted geometries. Figure S\ref{fig:ml_predictions_comparisons}(c) plots the Gibbs free energy of the three intermediates obtained with constraints (x-axis) and without constraints (y-axis). We find the intermediates relaxing to a lower energy minima once the constraints were lifted. A large number of data points considered exhibit dissociation and/or desorption when the constraints were lifted resulting in a large \gls{MAE} of 0.38 eV. A vast majority of these datapoints correspond to the \ce{OOH^*} intermediate. However, the Gibbs free energy of adsorption for dissociated or desorbed intermediates can not be appropriately interpreted due to the lack of adsorption. As such, we also evaluated the \gls{MAE} for intermediates where these events did not occur in both methods of predictions and demonstrated a more reasonable \gls{MAE} of 0.17 eV. Figure S\ref{fig:ml_predictions_comparisons}(d) shows the corresponding overpotentials interpreted from these two sets of Gibbs free energy and shows a similar behavior in \gls{MAE} when dissocition and desorption events are considered or omitted. Figure S\ref{fig:ml_predictions_comparisons}(e) shows a barplot distribution of the absolute difference in overpotential when considering the two methods of prediction. Although a significant amount of data points do exhibit dissociation and desorption with some overpotentials having inconsistent rate determining steps across the two methods, we see that these data points most correspond to larger disparities in overpotential with the majority of data points have a disparity of less than 0.25. 

\color{black}

\subsection{Database usage}

The entire database including the initial and relaxed structures and total energies are freely available through the University of Houston Dataverse Repository\cite{APJFTM_2023}. The database comes in 4,119 .json files\color{black}~(one for each material assessed) with the mpid name followed by the '.json' suffix (e.g. mp-775737.json). Each file contains a list\color{black}~of dictionary\color{black}~objects. Each dictionary contains the metadata and predicted information of a specific surface and all the surface intermediates (\ce{O^*}, \ce{OH^*}, and \ce{OOH^*}) of that surface and is structured as shown in Figure~S\ref{fig:dataframe}:

\begin{figure*}
    \centering
    \includegraphics[width=1\linewidth]{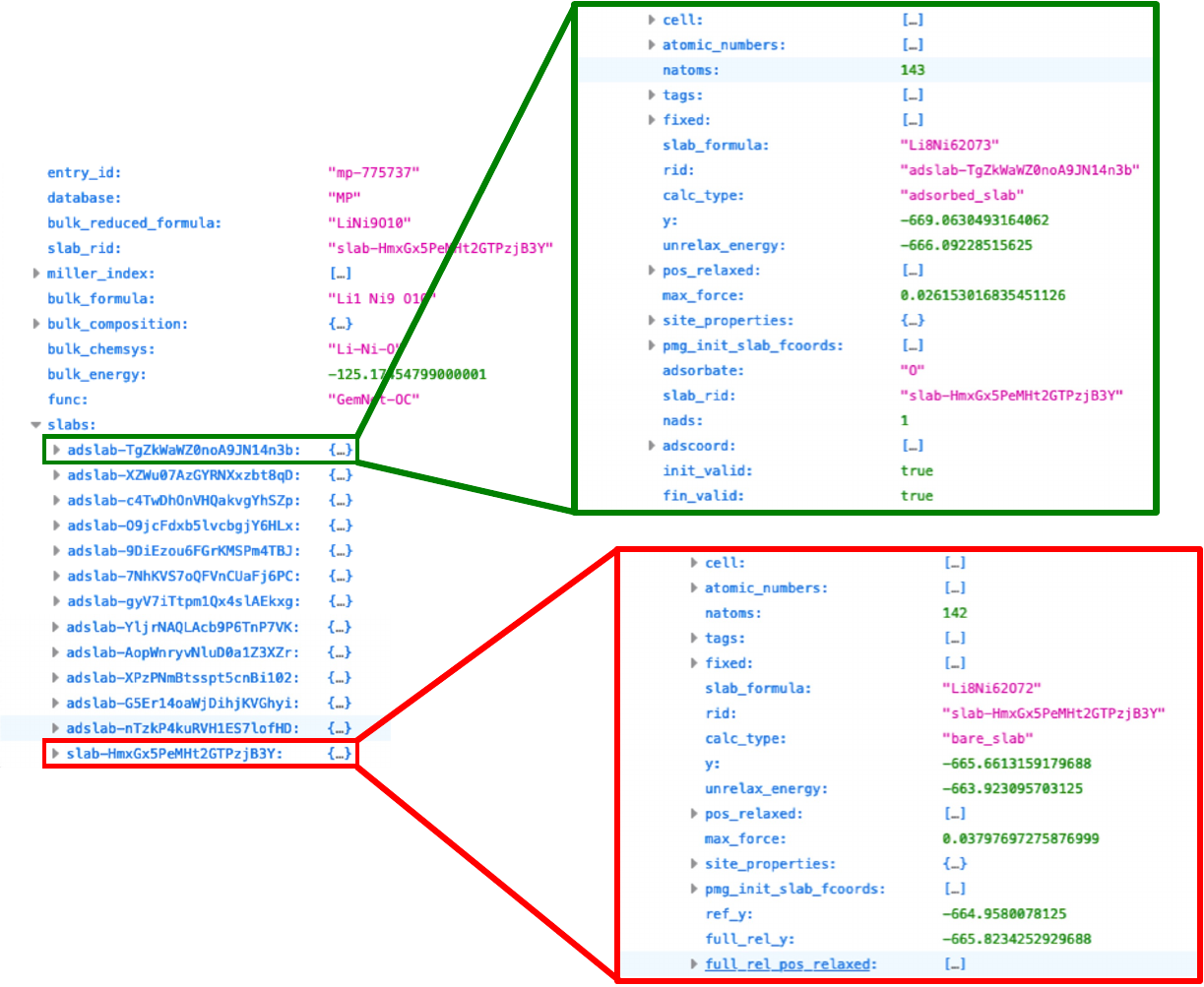}
    \caption{Dictionary object of a single entry in each database file.}
    \label{fig:dataframe}
\end{figure*}

Metadata of each surface includes the Materials Project ID (entry\_id\color{black}), the database from which the bulk structure used to generate the slab was obtained from (database\color{black}), the chemical formula of the bulk (bulk\_reduced\_formula\color{black}), a unique 20 character random ID that is assigned to all slabs considered (slab\_rid\color{black}) with the 'slab-' prefix indicating a bare slab structure, the Miller index (miller\_index\color{black}), the formula of the conventional unit cell (bulk\_formula\color{black}), a \textit{pymatgen} composition dictionary (bulk\_composition\color{black}), the chemical system (bulk\_chemsys\color{black}), the total energy of the \gls{DFT} computed conventional bulk structure (bulk\_energy\color{black}), the \gls{ML} architecture used to perform the predictions (func\color{black}), and a dictionary containing additional information pertaining to each individual adsorbed slab derived from the bare slab (slabs\color{black}).

The keys\color{black}~of the slabs dictionary are the unique 20 character random ID corresponding to each slab (designated with the 'slab-' prefix) and adsorbed slab (designated with the 'adslab-' prefix). Each value\color{black}~of the dictionary represents a single prediction of a slab or adsorbed slab and contains its predicted properties and metadata properties. These properties are the lattice parameter matrix of the slab (cell\color{black}), a list of atomic numbers making up the slab (atomic\_numbers\color{black}), the number of atoms in the slab (natoms\color{black}), tags indicating subsurface, surface and adsorbate atoms (tags\color{black}), the index of atoms that were fixed during relaxation (fixed\color{black}), the formula of the slab (slab\_formula\color{black}), the random ID (rid\color{black}), the type of calculation (calc\_type\color{black}), the final energy (y\color{black}), the unrelaxed energy (unrelax\_energy\color{black}), the reference slab energy used for calculating adsorption energy (ref\_y\color{black}) the relaxed xyz positions of all atoms in Cartesian coordinates (pos\_relaxed\color{black}), the maximum force during the final relaxation step (max\_force\color{black}), site properties such as the bulk Wyckoff positions (site\_properties\color{black}), and the fractional crystal coordinates of the initial structure (pmg\_init\_slab\_fcoords\color{black}). For entries corresponding to the adsorbed slabs, we also provided the random ID of the corresponding bare slab to help expedite binding energy calculations (slab\_rid\color{black}), the fractional coordinate position of the adsorbate (slab\_rid\color{black}) the intermediate (adsorbate\color{black}), and the number of adsorbates (nads\color{black}). As mentioned in the main manuscript, surfaces yielding low overpotentials of $\eta_{OER} < 0.75 V$ are also fully relaxed without the application of spring forces to maintain the intermediate identity an prevent desorption. These entries will have the relaxed energy (full\_rel\_y\color{black}) and relaxed coordinates (full\_rel\_pos\_relaxed\color{black}).

\clearpage
\section{Candidate materials}
\begin{table}[ht!]
\caption{\label{tab:best_facets_table_1} Summary of screening criteria for our final set of 190 candidate catalyst materials for \gls{OER} across all screening frameworks with the formula, space group, number of facets on the Wulff shape with $\eta < 0.75$ V, lowest overpotential across the facets, the screening framework used to identify this candidate (See superscript labels in Table 2 of the main manuscript), the Pourbaix formation energy ($E_{PBX}$), energy above hull ($E_{hull}$), and material cost. Entries are sorted by $E_{PBX}$.}

\begin{tabular}{c|c|c|c|c|c|c|c|c}
\hline
\hline
 & Space & \# & $\eta$ & Screening & $E_{PBX}$ & $E_{hull}$ & Cost \\
Formula & group & facets & (V) & framework & (eV) & (eV) & (\$/kg) \\
\hline
\hline
\ce{HgSeO4} & $Pmn2_1$ & 2 & 0.18 & a,d,e,h,i,l,m,p & 0.00 & 0.00 & 65.47 \\
\hline
\ce{Ni(BiO3)2} & $P4_2/mnm$ & 4 & 0.36 & a,b,c,d,e,f,g,h & 0.00 & 0.00 & 20.88 \\
 &  &  & & i,j,k,l,m,n,o,p & & & \\
\hline
\ce{Na2Se2O7} & $P\bar{1}$ & 2 & 0.21 & a,d,e,h,i,l,m,p & 0.00 & 0.00 & 110.43 \\
\hline
\ce{Ag3O4} & $P2_1/c$ & 4 & 0.33 & a,b,c,d,e,f,g,h & 0.00 & 0.00 & 714.77 \\
 &  &  & & i,j,k,l,m,n,o,p & & & \\
\hline
\ce{PbO2} & $P4_2/mnm$ & 2 & 0.33 & a,d,e,h,i,l,m,p & 0.00 & 0.00 & 2.41 \\
\hline
\ce{Mg(BiO3)2} & $P4_2/mnm$ & 3 & 0.52 & a,b,c,d,e,f,g,h & 0.00 & 0.00 & 20.41 \\
 &  &  & & i,j,k,l,m,n,o,p & & & \\
\hline
\ce{AgO} & $Cccm$ & 2 & 0.49 & a,d,e,h,i,l,m,p & 0.00 & 0.00 & 745.41 \\
\hline
\ce{AgO} & $C2/c$ & 2 & 0.51 & a,d,e,h,i,l,m,p & 0.01 & 0.01 & 745.41 \\
\hline
\ce{PbO2} & $Pbcn$ & 2 & 0.56 & a,b,c,d,e,f,g,h & 0.01 & 0.01 & 2.41 \\
 &  &  & & i,j,k,l,m,n,o,p & & & \\
\hline
\ce{Co(BiO3)2} & $P4_2/mnm$ & 2 & 0.33 & a,d,e,h,i,l,m,p & 0.02 & 0.02 & 24.34 \\
\hline
\ce{AgO} & $P2_1/c$ & 3 & 0.50 & a,b,c,d,e,f,g,h & 0.04 & 0.04 & 745.41 \\
 &  &  & & i,j,k,l,m,n,o,p & & & \\
\hline
\ce{Bi4O7} & $P\bar{1}$ & 2 & 0.22 & a,d,e,h,i,l,m,p & 0.04 & 0.00 & 22.67 \\
\hline
\ce{FeCo9O20} & $P\bar{1}$ & 4 & 0.41 & a,b,c,d,e,f,g,h & 0.06 & 0.07 & 31.63 \\
 &  &  & & i,j,k,l,m,n,o,p & & & \\
\hline
\ce{Ag2SeO4} & $Fddd$ & 2 & 0.40 & a,b,c,d,e,f,g,h & 0.07 & 0.00 & 547.23 \\
 &  &  & & i,j,k,l,m,n,o,p & & & \\
\hline
\ce{Bi3SbO7} & $P\bar{1}$ & 6 & 0.40 & a,b,c,d,e,f,g,h & 0.07 & 0.00 & 20.85 \\
 &  &  & & i,j,k,l,m,n,o,p & & & \\
\hline
\ce{CaBi4O9} & $P2/c$ & 2 & 0.37 & a,d,e,h,i,l,m,p & 0.07 & 0.00 & 21.36 \\
\hline
\ce{Li(CoO2)8} & $P\bar{1}$ & 4 & 0.46 & a,b,c,d,e,f,g,h & 0.08 & 0.03 & 34.60 \\
 &  &  & & i,j,k,l,m,n,o,p & & & \\
\hline
\ce{LiSbO3} & $Pnna$ & 2 & 0.61 & a,b,c,d,e,f,g,h & 0.09 & 0.00 & 10.82 \\
 &  &  & & i,j,k,l,m,n,o,p & & & \\
\hline
\ce{CoAgO3} & $R\bar{3}$ & 2 & 0.31 & a,b,c,d,e,f,g,h & 0.09 & 0.09 & 444.63 \\
 &  &  & & i,j,k,l,m,n,o,p & & & \\
\hline
\ce{AgSnO3} & $Cmmm$ & 2 & 0.49 & a,d,e,h,i,l,m,p & 0.09 & 0.09 & 350.79 \\
\hline
\hline
\end{tabular}
\end{table}

\begin{table}[ht!]
\caption{\label{tab:best_facets_table_2} Summary of screening criteria for our final set of candidates (continued).}

\begin{tabular}{c|c|c|c|c|c|c|c|c}
\hline
\hline
 & Space & \# & $\eta$ & Screening & $E_{PBX}$ & $E_{hull}$ & Cost \\
Formula & group & facets & (V) & framework & (eV) & (eV) & (\$/kg) \\
\hline
\hline
\ce{Na(CoO2)3} & $C2/m$ & 3 & 0.42 & a,b,c,d,e,f,g,h & 0.10 & 0.00 & 51.61 \\
 &  &  & & i,j,k,l,m,n,o,p & & & \\
\hline
\ce{Ce9YO20} & $P\bar{1}$ & 2 & 0.71 & a,d,e,h,i,l,m,p & 0.10 & 0.02 & 266.60 \\
\hline
\ce{Ce4SnO10} & $R\bar{3}m$ & 2 & 0.59 & a,b,c,d,e,f,g,h & 0.11 & 0.07 & 239.03 \\
 &  &  & & i,j,k,l,m,n,o,p & & & \\
\hline
\ce{Ag6Mo10O33} & $P\bar{1}$ & 3 & 0.55 & a,d,e,h,i,l,m,p & 0.11 & 0.02 & 278.69 \\
\hline
\ce{CdGe2O5} & $P\bar{1}$ & 2 & 0.42 & a,d,e,h,i,l,m,p & 0.11 & 0.01 & 625.43 \\
\hline
\ce{CuMoO4} & $P\bar{1}$ & 6 & 0.46 & a,b,c,d,e,f,g,h & 0.13 & 0.02 & 21.29 \\
 &  &  & & i,j,k,l,m,n,o,p & & & \\
\hline
\ce{ReAgO4} & $I4_1/a$ & 2 & 0.49 & a,d,e,h,i,l,m,p & 0.13 & 0.00 & 1737.20 \\
\hline
\ce{CuMoO4} & $P\bar{1}$ & 3 & 0.63 & a,b,c,d,e,f,g,h & 0.15 & 0.04 & 21.29 \\
 &  &  & & i,j,k,l,m,n,o,p & & & \\
\hline
\ce{Ce2Mo4O15} & $P\bar{1}$ & 4 & 0.46 & a,b,c,d,e,f,g,h & 0.16 & 0.02 & 126.84 \\
 &  &  & & i,j,k,l,m,n,o,p & & & \\
\hline
\ce{Ca(CuO2)2} & $Pbcm$ & 2 & 0.60 & a,b,c,d,e,f,g,h & 0.16 & 0.02 & 6.91 \\
 &  &  & & i,j,k,l,m,n,o,p & & & \\
\hline
\ce{Cu3Mo2O9} & $P\bar{1}$ & 2 & 0.55 & a,d,e,h,i,l,m,p & 0.18 & 0.03 & 19.31 \\
\hline
\ce{Ti2CoO5} & $Cmcm$ & 2 & 0.58 & a,b,c,d,e,f,g,h & 0.19 & 0.00 & 18.83 \\
 &  &  & & i,j,k,l,m,n,o,p & & & \\
\hline
\ce{TlCoO3} & $R\bar{3}$ & 2 & 0.41 & a,b,c,d,e,f,g,h & 0.19 & 0.05 & 3950.09 \\
 &  &  & & i,j,k,l,m,n,o,p & & & \\
\hline
\ce{Mn(SbO3)2} & $P321$ & 2 & 0.48 & a,b,c,d,e,f,g,h & 0.20 & 0.00 & 9.84 \\
 &  &  & & i,j,k,l,m,n,o,p & & & \\
\hline
\ce{CuWO4} & $P\bar{1}$ & 2 & 0.39 & a,b,c,d,e,f,g,h & 0.20 & 0.08 & 23.40 \\
 &  &  & & i,j,k,l,m,n,o,p & & & \\
\hline
\ce{NiBiO3} & $Pnma$ & 2 & 0.50 & a,b,c,d,e,f,g,h & 0.21 & 0.04 & 20.66 \\
 &  &  & & i,j,k,l,m,n,o,p & & & \\
\hline
\ce{Fe(Bi5O8)5} & $P23$ & 2 & 0.70 & a,d,e,h,i,l,m,p & 0.21 & 0.03 & 22.66 \\
\hline
\ce{Cu3(SbO3)4} & $Im\bar{3}$ & 2 & 0.37 & a,d,e,h,i,l,m,p & 0.21 & 0.02 & 10.76 \\
\hline
\ce{Bi2O3} & $P2_1/c$ & 2 & 0.52 & a,d,e,h,i,l,m,p & 0.22 & 0.00 & 23.01 \\
\hline
\ce{Cr2Ag2O7} & $P\bar{1}$ & 3 & 0.52 & a,b,c,d,e,f,g,h & 0.22 & 0.02 & 430.56 \\
 &  &  & & i,j,k,l,m,n,o,p & & & \\
\hline
\ce{Cr2Ag2O7} & $P\bar{1}$ & 3 & 0.43 & a,b,c,d,e,f,g,h & 0.22 & 0.03 & 430.56 \\
 &  &  & & i,j,k,l,m,n,o,p & & & \\
\hline
\ce{AlTlO3} & $Pnma$ & 2 & 0.55 & a,b,c,d,e,f,g,h & 0.22 & 0.08 & 4391.10 \\
 &  &  & & i,j,k,l,m,n,o,p & & & \\
\hline
\ce{LiAgO2} & $C2/m$ & 2 & 0.39 & a,d,e,h,i,l,m,p & 0.22 & 0.02 & 629.44 \\
\hline
\hline
\end{tabular}
\end{table}

\begin{table}[ht!]
\caption{\label{tab:best_facets_table_3} Summary of screening criteria for our final set of candidates (continued).}

\begin{tabular}{c|c|c|c|c|c|c|c|c}
\hline
\hline
 & Space & \# & $\eta$ & Screening & $E_{PBX}$ & $E_{hull}$ & Cost \\
Formula & group & facets & (V) & framework & (eV) & (eV) & (\$/kg) \\
\hline
\hline
\ce{Nb2Cu3O8} & $P\bar{1}$ & 3 & 0.36 & a,b,c,d,e,f,g,h & 0.23 & -7.61 & 19.94 \\
 &  &  & & i,j,k,l,m,n,o,p & & & \\
\hline
\ce{Cd(CoO2)2} & $Pmmn$ & 2 & 0.43 & a,b,c,d,e,f,g,h & 0.23 & -6.26 & 22.61 \\
 &  &  & & i,j,k,l,m,n,o,p & & & \\
\hline
\ce{Bi2O3} & $Pbcn$ & 3 & 0.40 & a,b,c,d,e,f,g,h & 0.24 & 0.02 & 23.01 \\
 &  &  & & i,j,k,l,m,n,o,p & & & \\
\hline
\ce{Ag2BiO3} & $Pnna$ & 3 & 0.34 & a,b,c,d,e,f,g,h & 0.24 & 0.00 & 401.94 \\
 &  &  & & i,j,k,l,m,n,o,p & & & \\
\hline
\ce{K(CoO2)2} & $P2_1$ & 3 & 0.51 & a,b,c,d,e,f,g,h & 0.24 & 0.00 & 205.67 \\
 &  &  & & i,j,k,l,m,n,o,p & & & \\
\hline
\ce{ZnCoO3} & $C2/c$ & 2 & 0.49 & a,d,e,h,i,l,m,p & 0.25 & 0.06 & 19.81 \\
\hline
\ce{Li(Bi3O5)4} & $I23$ & 2 & 0.47 & a,d,e,h,i,l,m,p & 0.25 & 0.08 & 22.74 \\
\hline
\ce{CoAgO2} & $P6_3/mmc$ & 2 & 0.47 & a,b,c,d,e,f,g,h & 0.26 & 0.00 & 480.17 \\
 &  &  & & i,j,k,l,m,n,o,p & & & \\
\hline
\ce{Cu(BiO2)2} & $P4/ncc$ & 3 & 0.41 & a,b,c,d,e,f,g,h & 0.26 & 0.01 & 20.85 \\
 &  &  & & i,j,k,l,m,n,o,p & & & \\
\hline
\ce{CdCoO3} & $C2/c$ & 2 & 0.46 & a,d,e,h,i,l,m,p & 0.28 & 0.05 & 16.08 \\
\hline
\ce{Cd2PbO4} & $Pbam$ & 3 & 0.22 & a,b,c,d,e,f,g,h & 0.29 & 0.00 & 2.59 \\
 &  &  & & i,j,k,l,m,n,o,p & & & \\
\hline
\ce{Ag2GeO3} & $P2_12_12_1$ & 2 & 0.62 & a,d,e,h,i,l,m,p & 0.29 & 0.00 & 862.25 \\
\hline
\ce{MnMoO5} & $P\bar{1}$ & 2 & 0.56 & a,d,e,h,i,l,m,p & 0.30 & -8.27 & 18.68 \\
\hline
\ce{ScCoO3} & $P2_1/c$ & 2 & 0.34 & a,b,c,d,e,f,g,h & 0.30 & 0.06 & 1045.31 \\
 &  &  & & i,j,k,l,m,n,o,p & & & \\
\hline
\ce{Mn(SeO3)2} & $P2_1/c$ & 4 & 0.35 & a,b,c,d,e,f,g,h & 0.31 & 0.00 & 75.94 \\
 &  &  & & i,j,k,l,m,n,o,p & & & \\
\hline
\ce{Cr3AgO8} & $C2/m$ & 2 & 0.62 & a,b,c,d,e,f,g,h & 0.32 & 0.00 & 240.23 \\
 &  &  & & i,j,k,l,m,n,o,p & & & \\
\hline
\ce{InCoO3} & $Pnma$ & 2 & 0.58 & a,b,c,d,e,f,g,h & 0.32 & 0.07 & 387.34 \\
 &  &  & & i,j,k,l,m,n,o,p & & & \\
\hline
\ce{Ca(FeO2)2} & $Pm$ & 2 & 0.59 & a,d,e,h,i,l,m,p & 0.33 & 0.01 & 1.90 \\
\hline
\ce{MnO2} & $Pnma$ & 2 & 0.41 & a,b,c,d,i,j,k,l & 0.33 & 0.00 & 2.41 \\
\hline
\ce{TiAg2O3} & $C2/c$ & 2 & 0.47 & a,d,e,h,i,l,m,p & 0.33 & 0.04 & 594.55 \\
\hline
\ce{CoPbO3} & $R\bar{3}$ & 3 & 0.36 & a,d,e,h,i,l,m,p & 0.34 & 0.03 & 11.78 \\
\hline
\ce{CaBiO3} & $P2_1/c$ & 3 & 0.35 & a,b,c,d,e,f,g,h & 0.34 & 0.00 & 18.94 \\
 &  &  & & i,j,k,l,m,n,o,p & & & \\
\hline
\ce{Tl3Co3O8} & $P1$ & 2 & 0.50 & a,d,e,h,i,l,m,p & 0.35 & 0.07 & 4018.88 \\
\hline
\hline
\end{tabular}
\end{table}

\begin{table}[ht!]
\caption{\label{tab:best_facets_table_4} Summary of screening criteria for our final set of candidates (continued).}

\begin{tabular}{c|c|c|c|c|c|c|c|c}
\hline
\hline
 & Space & \# & $\eta$ & Screening & $E_{PBX}$ & $E_{hull}$ & Cost \\
Formula & group & facets & (V) & framework & (eV) & (eV) & (\$/kg) \\
\hline
\hline
\ce{SrBiO3} & $P2_1/c$ & 3 & 0.39 & a,b,c,d,e,f,g,h & 0.35 & 0.00 & 16.06 \\
 &  &  & & i,j,k,l,m,n,o,p & & & \\
\hline
\ce{Ce11O20} & $P\bar{1}$ & 4 & 0.32 & a,b,c,d,e,f,g,h & 0.35 & 0.00 & 290.42 \\
 &  &  & & i,j,k,l,m,n,o,p & & & \\
\hline
\ce{NaBi5O8} & $P\bar{1}$ & 3 & 0.52 & a,b,c,d,e,f,g,h & 0.35 & 0.03 & 27.24 \\
 &  &  & & i,j,k,l,m,n,o,p & & & \\
\hline
\ce{TlBiO4} & $Cmmm$ & 2 & 0.33 & a,d,i,l & 0.36 & -5.53 & 2580.81 \\
\hline
\ce{Cu4Se3O10} & $P\bar{1}$ & 2 & 0.52 & a,d,e,h,i,l,m,p & 0.36 & 0.02 & 57.57 \\
\hline
\ce{BaBiO3} & $P2_1/c$ & 2 & 0.63 & a,d,e,h,i,l,m,p & 0.36 & 0.00 & 13.87 \\
\hline
\ce{Ce2(GeO3)3} & $P\bar{1}$ & 2 & 0.55 & a,b,c,d,e,f,g,h & 0.36 & 0.05 & 645.52 \\
 &  &  & & i,j,k,l,m,n,o,p & & & \\
\hline
\ce{CdSe2O5} & $C2/c$ & 2 & 0.66 & a,d,e,h,i,l,m,p & 0.36 & 0.00 & 67.37 \\
\hline
\ce{AgBiO2} & $P2_1/m$ & 3 & 0.55 & a,b,c,d,e,f,g,h & 0.36 & 0.00 & 279.98 \\
 &  &  & & i,j,k,l,m,n,o,p & & & \\
\hline
\ce{Zn(BiO2)2} & $P\bar{1}$ & 3 & 0.43 & a,b,c,d,e,f,g,h & 0.36 & 0.07 & 20.03 \\
 &  &  & & i,j,k,l,m,n,o,p & & & \\
\hline
\ce{CuSeO3} & $P2_1/c$ & 4 & 0.41 & a,b,c,d,e,f,g,h & 0.37 & 0.03 & 64.44 \\
 &  &  & & i,j,k,l,m,n,o,p & & & \\
\hline
\ce{Cu2SeO4} & $P2_1/c$ & 4 & 0.38 & a,b,c,d,e,f,g,h & 0.37 & 0.03 & 47.88 \\
 &  &  & & i,j,k,l,m,n,o,p & & & \\
\hline
\ce{CdSeO3} & $Pnma$ & 4 & 0.49 & a,b,c,d,e,f,g,h & 0.37 & 0.01 & 50.04 \\
 &  &  & & i,j,k,l,m,n,o,p & & & \\
\hline
\ce{FeSnO3} & $P\bar{1}$ & 2 & 0.51 & a,b,c,d,e,f,g,h & 0.38 & 0.00 & 18.17 \\
 &  &  & & i,j,k,l,m,n,o,p & & & \\
\hline
\ce{CoGeO3} & $C2/c$ & 3 & 0.50 & a,b,c,d,e,f,g,h & 0.38 & 0.06 & 604.48 \\
 &  &  & & i,j,k,l,m,n,o,p & & & \\
\hline
\ce{Hg2MoO4} & $C2/c$ & 2 & 0.26 & a,d,e,h,i,l,m,p & 0.38 & 0.01 & 45.80 \\
\hline
\ce{CdIn2O4} & $Imma$ & 2 & 0.46 & a,b,c,d,e,f,g,h & 0.38 & 0.08 & 408.44 \\
 &  &  & & i,j,k,l,m,n,o,p & & & \\
\hline
\ce{BaTl2O4} & $Pnma$ & 3 & 0.54 & a,b,c,d,e,f,g,h & 0.38 & 0.00 & 4021.08 \\
 &  &  & & i,j,k,l,m,n,o,p & & & \\
\hline
\ce{Cr(SbO3)2} & $Pnnm$ & 2 & 0.39 & a,b,c,d,e,f,g,h & 0.38 & 0.06 & 10.88 \\
 &  &  & & i,j,k,l,m,n,o,p & & & \\
\hline
\ce{Li3BiO4} & $P2/c$ & 2 & 0.41 & a,b,c,d,i,j,k,l & 0.38 & 0.00 & 18.93 \\
\hline
\ce{SrSe2O5} & $P\bar{1}$ & 3 & 0.25 & a,b,c,d,e,f,g,h & 0.39 & 0.00 & 71.87 \\
 &  &  & & i,j,k,l,m,n,o,p & & & \\
\hline
\ce{Ca2Se3O8} & $P\bar{1}$ & 2 & 0.70 & a,d,e,h,i,l,m,p & 0.39 & 0.00 & 79.44 \\
\hline
\hline
\end{tabular}
\end{table}

\begin{table}[ht!]
\caption{\label{tab:best_facets_table_5} Summary of screening criteria for our final set of candidates (continued).}

\begin{tabular}{c|c|c|c|c|c|c|c|c}
\hline
\hline
 & Space & \# & $\eta$ & Screening & $E_{PBX}$ & $E_{hull}$ & Cost \\
Formula & group & facets & (V) & framework & (eV) & (eV) & (\$/kg) \\
\hline
\hline
\ce{Cd6(CoO3)5} & $R32$ & 2 & 0.39 & a,b,c,d,e,f,g,h & 0.39 & 0.09 & 14.84 \\
 &  &  & & i,j,k,l,m,n,o,p & & & \\
\hline
\ce{Ag2SeO3} & $P2_1/c$ & 3 & 0.29 & a,b,c,d,g,h,i,j & 0.40 & 0.00 & 572.64 \\
 &  &  & & k,l,o,p & & & \\
\hline
\ce{BaSe2O5} & $P2_1/c$ & 2 & 0.66 & a,d,e,h,i,l,m,p & 0.40 & 0.00 & 62.17 \\
\hline
\ce{LuCoO3} & $Pnma$ & 2 & 0.56 & a,b,c,d,e,f,g,h & 0.40 & 0.03 & 4666.55 \\
 &  &  & & i,j,k,l,m,n,o,p & & & \\
\hline
\ce{Ag4GeO4} & $P\bar{1}$ & 6 & 0.52 & a,b,c,d,e,f,g,h & 0.40 & 0.01 & 835.50 \\
 &  &  & & i,j,k,l,m,n,o,p & & & \\
\hline
\ce{CuReO4} & $P\bar{1}$ & 4 & 0.33 & a,b,c,d,e,f,g,h & 0.40 & 0.06 & 1690.40 \\
 &  &  & & i,j,k,l,m,n,o,p & & & \\
\hline
\ce{Co11CuO16} & $P2/m$ & 2 & 0.56 & a,d,i,l & 0.40 & 0.06 & 36.39 \\
\hline
\ce{CoSe2O5} & $Pbcn$ & 2 & 0.40 & a,b,c,d & 0.41 & 0.00 & 88.84 \\
\hline
\ce{Co5SbO8} & $R\bar{3}m$ & 2 & 0.49 & a,b,c,d,i,j,k,l & 0.41 & 0.00 & 32.16 \\
\hline
\ce{NaTlO2} & $I4_1/amd$ & 2 & 0.49 & a,d,e,h,i,l,m,p & 0.41 & 0.02 & 4751.26 \\
\hline
\ce{MgIn2O4} & $Imma$ & 2 & 0.42 & a,b,c,d,e,f,g,h & 0.41 & 0.02 & 521.03 \\
 &  &  & & i,j,k,l,m,n,o,p & & & \\
\hline
\ce{Li7Co5O12} & $C2/m$ & 4 & 0.38 & a,b,c,d,e,f,g,h & 0.41 & 0.03 & 30.17 \\
 &  &  & & i,j,k,l,m,n,o,p & & & \\
\hline
\ce{Hg2WO4} & $C2/c$ & 2 & 0.68 & a,d,e,h,i,l,m,p & 0.42 & 0.00 & 43.49 \\
\hline
\ce{CuReO4} & $P\bar{1}$ & 3 & 0.56 & a,b,c,d,e,f,g,h & 0.42 & 0.08 & 1690.40 \\
 &  &  & & i,j,k,l,m,n,o,p & & & \\
\hline
\ce{Sn2Ge2O7} & $P\bar{1}$ & 2 & 0.55 & a,d,e,h,i,l,m,p & 0.42 & 0.06 & 442.22 \\
\hline
\ce{MnTlO3} & $Pnma$ & 5 & 0.20 & a,b,c,d,i,j,k,l & 0.42 & 0.05 & 3991.79 \\
\hline
\ce{Ag2HgO2} & $P4_32_12$ & 3 & 0.47 & a,b,c,d,g,h,i,j & 0.43 & 0.00 & 435.93 \\
 &  &  & & k,l,o,p & & & \\
\hline
\ce{Co2NiO4} & $Imma$ & 2 & 0.50 & a,b,c,d,e,f,g,h & 0.43 & 0.09 & 30.91 \\
 &  &  & & i,j,k,l,m,n,o,p & & & \\
\hline
\ce{Ba8(Bi2O7)3} & $P\bar{1}$ & 6 & 0.46 & a,b,c,d,i,j,k,l & 0.43 & 0.00 & 12.29 \\
\hline
\ce{AgSbO4} & $Cmmm$ & 3 & 0.32 & a,b,c,d,i,j,k,l & 0.44 & -5.61 & 320.88 \\
\hline
\ce{MnTlO3} & $P\bar{1}$ & 3 & 0.36 & a,b,c,d,i,j,k,l & 0.44 & 0.07 & 3991.79 \\
\hline
\ce{Ca(CoO2)2} & $Pnma$ & 2 & 0.32 & a,b,c,d,e,f,g,h & 0.44 & 0.00 & 29.47 \\
 &  &  & & i,j,k,l,m,n,o,p & & & \\
\hline
\ce{Tl2SeO4} & $Pnma$ & 3 & 0.58 & a,b,c,d,i,j,k,l & 0.45 & 0.00 & 4467.30 \\
\hline
\ce{Tl2SeO4} & $P2_12_12_1$ & 4 & 0.52 & a,b,c,d,i,j,k,l & 0.45 & 0.00 & 4467.30 \\
\hline
\ce{CaSeO3} & $Pnma$ & 2 & 0.60 & a,b,c,d,g,h,i,j & 0.45 & 0.01 & 71.04 \\
 &  &  & & k,l,o,p & & & \\
\hline
\hline
\end{tabular}
\end{table}

\begin{table}[ht!]
\caption{\label{tab:best_facets_table_6} Summary of screening criteria for our final set of candidates (continued).}

\begin{tabular}{c|c|c|c|c|c|c|c|c}
\hline
\hline
 & Space & \# & $\eta$ & Screening & $E_{PBX}$ & $E_{hull}$ & Cost \\
Formula & group & facets & (V) & framework & (eV) & (eV) & (\$/kg) \\
\hline
\hline
\ce{CdCu2O3} & $Pmmn$ & 2 & 0.50 & a,b,c,d,g,h,i,j & 0.46 & 0.04 & 5.78 \\
 &  &  & & k,l,o,p & & & \\
\hline
\ce{CuSbO4} & $Cmmm$ & 2 & 0.36 & a,b,c,d,i,j,k,l & 0.47 & -6.01 & 10.18 \\
\hline
\ce{YMn2O5} & $Pbam$ & 2 & 0.54 & a,b,c,d,i,j,k,l & 0.47 & 0.00 & 11.56 \\
\hline
\ce{Cu2W2O7} & $P\bar{1}$ & 2 & 0.43 & a,b,c,d,i,j,k,l & 0.47 & 0.10 & 23.94 \\
\hline
\ce{Ca3WO6} & $P2_1/c$ & 2 & 0.49 & a,d,e,h,i,l,m,p & 0.47 & 0.02 & 18.40 \\
\hline
\ce{YCoO3} & $P2_1/c$ & 2 & 0.52 & a,b,c,d,i,j,k,l & 0.48 & 0.01 & 30.52 \\
\hline
\ce{ZnCu2O3} & $Pmmn$ & 3 & 0.42 & a,b,c,d,g,h,i,j & 0.48 & 0.10 & 6.44 \\
 &  &  & & k,l,o,p & & & \\
\hline
\ce{CuNiO2} & $C2/m$ & 2 & 0.40 & a,b,c,d,i,j,k,l & 0.49 & 0.07 & 11.60 \\
\hline
\ce{MnBiO3} & $Pnma$ & 3 & 0.51 & a,b,c,d,k,l & 0.50 & 0.03 & 17.78 \\
\hline
\hline
\end{tabular}
\end{table}

\begin{table}[ht!]
\caption{\label{tab:best_facets_table_7} Summary of screening criteria for our final set of candidates (continued). All materials listed from here are unstable as bulk materials ($E_{PBX} > 0.5$ eV) but can be stabilized as nanoparticles.}

\begin{tabular}{c|c|c|c|c|c|c|c|c}
\hline
\hline
 & Space & \# & $\eta$ & Screening & $E_{PBX}$ & $E_{hull}$ & Cost \\
Formula & group & facets & (V) & framework & (eV) & (eV) & (\$/kg) \\
\hline
\hline
\ce{Cr2WO6} & $P4_2/mnm$ & 2 & 0.25 & c,d,k,l & 0.52 & 0.00 & 20.20 \\
\hline
\ce{TlCuO2} & $R\bar{3}m$ & 2 & 0.43 & h,p & 0.52 & 0.05 & 4091.65 \\
\hline
\ce{Y(FeO2)2} & $P\bar{1}$ & 2 & 0.53 & c,d,g,h,k,l,o,p & 0.54 & 0.01 & 11.23 \\
\hline
\ce{Sr2Tl2O5} & $P2_1/c$ & 2 & 0.37 & k,l & 0.55 & 0.00 & 3694.92 \\
\hline
\ce{ZrCoO3} & $P\bar{1}$ & 3 & 0.46 & c,d,g,h,k,l,o,p & 0.55 & 0.10 & 32.83 \\
\hline
\ce{TiVO4} & $P2_1$ & 2 & 0.58 & c,d,k,l & 0.56 & 0.02 & 120.33 \\
\hline
\ce{HfFeO3} & $Pnma$ & 2 & 0.53 & c,d,k,l & 0.56 & 0.06 & 569.53 \\
\hline
\ce{Mn4CuO8} & $C2/m$ & 3 & 0.39 & d,l & 0.56 & 0.07 & 3.51 \\
\hline
\ce{CoCu2O3} & $Pmmn$ & 3 & 0.42 & c,d,k,l & 0.57 & 0.07 & 18.94 \\
\hline
\ce{MnSe2O5} & $Pbcn$ & 2 & 0.50 & c,d,k,l & 0.58 & 0.00 & 79.93 \\
\hline
\ce{CrMoO4} & $Cmmm$ & 2 & 0.42 & c,d,k,l & 0.59 & 0.00 & 21.89 \\
\hline
\ce{KMn2O4} & $P\bar{1}$ & 2 & 0.47 & d,l & 0.60 & 0.00 & 185.55 \\
\hline
\ce{Ba2Tl2O5} & $Pnma$ & 2 & 0.34 & d,l & 0.60 & 0.00 & 3213.59 \\
\hline
\ce{LuMnO3} & $Pnma$ & 3 & 0.54 & c,d,g,h,k,l,o,p & 0.61 & 0.05 & 4722.98 \\
\hline
\ce{TiMnO3} & $R\bar{3}$ & 2 & 0.28 & d,l & 0.62 & 0.00 & 5.36 \\
\hline
\ce{KBiO2} & $C2/c$ & 2 & 0.27 & c,d,k,l & 0.62 & 0.00 & 158.83 \\
\hline
\ce{Na5ReO6} & $C2/m$ & 2 & 0.30 & h,l,p & 0.62 & 0.00 & 1406.49 \\
\hline
\ce{CuTeO4} & $Cmmm$ & 3 & 0.54 & c,d,k,l & 0.63 & -5.71 & 177.66 \\
\hline
\ce{ScCrO3} & $Pnma$ & 2 & 0.46 & c,d,g,h,k,l,o,p & 0.63 & 0.04 & 1077.47 \\
\hline
\ce{Ta2CrO6} & $P4_2/mnm$ & 2 & 0.56 & c,d,k,l & 0.64 & 0.01 & 109.26 \\
\hline
\ce{MnSnO3} & $R\bar{3}$ & 2 & 0.43 & c,d,k,l & 0.65 & 0.00 & 18.71 \\
\hline
\ce{Li4PbO4} & $Cmcm$ & 2 & 0.66 & d,l & 0.65 & 0.00 & 2.61 \\
\hline
\ce{VSbO4} & $Cmmm$ & 2 & 0.44 & d,h,l,p & 0.66 & 0.02 & 87.80 \\
\hline
\ce{ScCuO2} & $R\bar{3}m$ & 2 & 0.41 & d,h,l,p & 0.66 & 0.00 & 1112.09 \\
\hline
\ce{Mn2BeO4} & $Pnma$ & 3 & 0.32 & c,d,g,h,k,l,o,p & 0.66 & 0.04 & 24.36 \\
\hline
\ce{AlCuO2} & $P6_3/mmc$ & 2 & 0.48 & c,d,k,l,o,p & 0.67 & 0.00 & 6.28 \\
\hline
\ce{TiCu3O4} & $P2_1/c$ & 2 & 0.38 & g,h,o,p & 0.68 & 0.07 & 8.46 \\
\hline
\ce{Ag2PbO2} & $C2/c$ & 3 & 0.43 & g,h,o,p & 0.70 & 0.00 & 406.97 \\
\hline
\ce{LuCrO3} & $Pnma$ & 2 & 0.43 & c,d,g,h,k,l,o,p & 0.71 & 0.00 & 4774.87 \\
\hline
\ce{VSeO4} & $P2_1/c$ & 2 & 0.46 & d,l & 0.72 & 0.00 & 157.65 \\
\hline
\ce{Ag3RuO4} & $P4_122$ & 2 & 0.60 & d,l & 0.72 & 0.07 & 5554.14 \\
\hline
\ce{GePb5O7} & $Pbca$ & 2 & 0.68 & g,h,k,l,o,p & 0.72 & 0.01 & 88.53 \\
\hline
\ce{VCrO4} & $Cmcm$ & 2 & 0.32 & c,d & 0.73 & 0.01 & 116.99 \\
\hline
\ce{TlTeO4} & $Cmmm$ & 2 & 0.50 & c,d,k,l & 0.74 & -5.50 & 3210.31 \\
\hline
\ce{MnSeO3} & $P2_1/c$ & 2 & 0.44 & c,d,k,l & 0.75 & 0.00 & 64.79 \\
\hline
\ce{Li3BiO3} & $P\bar{1}$ & 2 & 0.52 & d,h,l,p & 0.75 & 0.00 & 19.85 \\
\hline
\ce{VZn2O4} & $Imma$ & 2 & 0.47 & k,l & 0.75 & 0.02 & 79.05 \\
\hline
\ce{Fe10O11} & $P\bar{1}$ & 3 & 0.50 & g,h,o,p & 0.77 & 0.05 & 0.88 \\
\hline
\hline
\end{tabular}
\end{table}

\begin{table}[ht!]
\caption{\label{tab:best_facets_table_8} Summary of screening criteria for our final set of candidates (continued).}

\begin{tabular}{c|c|c|c|c|c|c|c|c}
\hline
\hline
 & Space & \# & $\eta$ & Screening & $E_{PBX}$ & $E_{hull}$ & Cost \\
Formula & group & facets & (V) & framework & (eV) & (eV) & (\$/kg) \\
\hline
\hline
\ce{Tl2SnO3} & $Pnma$ & 2 & 0.60 & d,h & 0.80 & 0.00 & 4269.57 \\
\hline
\ce{ScMn2O4} & $C2/m$ & 3 & 0.29 & k,l & 0.84 & 0.03 & 712.73 \\
\hline
\ce{ZrMnO3} & $R\bar{3}$ & 2 & 0.50 & c,d,g,h,k,l,o,p & 0.84 & 0.03 & 18.24 \\
\hline
\ce{Li(CuO)2} & $Pnma$ & 2 & 0.52 & k,l & 0.85 & 0.00 & 8.03 \\
\hline
\ce{Fe17O18} & $P\bar{1}$ & 2 & 0.52 & l & 0.86 & 0.04 & 0.86 \\
\hline
\ce{K6Co2O7} & $P2_1/c$ & 4 & 0.36 & c,d,k,l & 0.87 & 0.00 & 519.06 \\
\hline
\ce{Cr2NiO4} & $P1$ & 3 & 0.38 & d,h,l,p & 0.88 & 0.04 & 9.96 \\
\hline
\ce{YCuO2} & $P6_3/mmc$ & 2 & 0.38 & k,l & 0.94 & 0.00 & 18.75 \\
\hline
\ce{Mn2SnO4} & $Imma$ & 3 & 0.42 & c,d,g,h,k,l,o,p & 0.94 & 0.00 & 14.72 \\
\hline
\ce{MnCuO2} & $P\bar{1}$ & 2 & 0.58 & l & 0.94 & 0.00 & 5.41 \\
\hline
\ce{YCuO2} & $R\bar{3}m$ & 2 & 0.38 & g,h,o,p & 0.94 & 0.00 & 18.75 \\
\hline
\ce{Mn23FeO32} & $P\bar{1}$ & 2 & 0.28 & d,h,l,p & 0.95 & 0.01 & 2.27 \\
\hline
\ce{SrCr2O4} & $Pmmn$ & 3 & 0.62 & c,d,k,l & 0.98 & 0.00 & 4.97 \\
\hline
\ce{KSnO2} & $P\bar{1}$ & 2 & 0.19 & c,d,g,h,k,l,o,p & 0.98 & 0.02 & 226.98 \\
\hline
\ce{CdRuO4} & $Cmmm$ & 2 & 0.53 & d,l & 0.99 & -6.31 & 8784.84 \\
\hline
\ce{Co(SbO2)2} & $P4_2/mbc$ & 2 & 0.66 & k,l & 0.99 & 0.00 & 18.42 \\
\hline
\ce{Mn2CrO4} & $Cc$ & 2 & 0.37 & c,d,k,l & 1.04 & 0.00 & 4.02 \\
\hline
\ce{Na2Sb4O7} & $C2/c$ & 2 & 0.35 & k,l & 1.05 & 0.00 & 29.14 \\
\hline
\ce{CeCrO3} & $Pnma$ & 2 & 0.29 & c,d,g,h,k,l,o,p & 1.05 & 0.05 & 206.93 \\
\hline
\ce{Na2Co2O3} & $P2_1/c$ & 4 & 0.30 & g,h,o,p & 1.07 & 0.00 & 83.99 \\
\hline
\ce{NaSb5O8} & $P\bar{1}$ & 3 & 0.52 & k,l & 1.09 & 0.00 & 19.53 \\
\hline
\ce{RuPbO4} & $Cmmm$ & 2 & 0.64 & d,l & 1.12 & -6.84 & 6548.46 \\
\hline
\ce{SnRuO4} & $Cmmm$ & 2 & 0.32 & c,d,k,l & 1.14 & -7.30 & 8602.47 \\
\hline
\ce{Ti(SnO2)2} & $P4_2/mbc$ & 2 & 0.48 & k,l & 1.15 & 0.00 & 24.39 \\
\hline
\ce{LiMn3O4} & $P\bar{1}$ & 2 & 0.63 & d,l & 1.26 & 0.02 & 2.37 \\
\hline
\ce{K2PbO2} & $P\bar{1}$ & 2 & 0.55 & k,l & 1.31 & 0.00 & 248.19 \\
\hline
\ce{Mn2NiO3} & $Immm$ & 2 & 0.47 & d,h,l,p & 1.34 & 0.08 & 6.73 \\
\hline
\ce{Mn3NiO4} & $Cmmm$ & 2 & 0.54 & d,l & 1.44 & 0.09 & 5.63 \\
\hline
\ce{BaMn2O3} & $Immm$ & 2 & 0.62 & k,l & 1.66 & 0.00 & 1.38 \\
\hline
\ce{CaMn7O8} & $C2/m$ & 2 & 0.48 & l & 1.66 & 0.03 & 2.48 \\
\hline
\hline
\end{tabular}
\end{table}

\clearpage

\section{Candidate materials literature references}
\begin{table}[ht!]
\caption{\label{tab:best_facets_exp_comp_1}  Overpotentials from \gls{OC22}, the experimental literature, and \gls{DFT} (see Figure 7(b) in the main manuscript) along with the \gls{PDS} for our final set of 190 candidate catalyst materials for OER across all screening frameworks. Experimental results are reported for systems containing similar chemical systems and do not necessarily reflect the same formula of the candidate catalyst. Candidates are listed in the same order as Tables S\ref{tab:best_facets_table_1}-\ref{tab:best_facets_table_8} (from lowest to highest $E_{PBX}$).}

\begin{tabular}{c|c|c|c|c}
\hline
\hline
 & $\eta$ (V) &  $\eta$ (V) &  $\eta$ (V) &  \\
Formula & (OC22) & (exp.) & (DFT) & PDS \\
\hline
\hline
\ce{HgSeO4} & 0.176 &  & 0.217 & \ce{OH -> O^*+H^+} \\
\hline
\ce{Ni(BiO3)2} & 0.361 & 0.300\cite{Yu2022} &  & \ce{OH -> O^*+H^+} \\
\hline
\ce{Na2Se2O7} & 0.208 &  & 0.741 & \ce{OH -> O^*+H^+} \\
\hline
\ce{Ag3O4} & 0.333 & 0.37\cite{Joya2016} &  & \ce{OH -> O^*+H^+} \\
\hline
\ce{PbO2} & 0.334 &  &  & \ce{OH -> O^*+H^+} \\
\hline
\ce{Mg(BiO3)2} & 0.525 &  &  & \ce{OH -> O^*+H^+} \\
\hline
\ce{AgO} & 0.495 & 0.37\cite{Joya2016} &  & \ce{H2O+O^* -> OOH^*+H^+} \\
\hline
\ce{AgO} & 0.513 & 0.37\cite{Joya2016} &  & \ce{H2O+O^* -> OOH^*+H^+} \\
\hline
\ce{PbO2} & 0.561 &  &  & \ce{H2O+O^* -> OOH^*+H^+} \\
\hline
\ce{Co(BiO3)2} & 0.331 & 0.320\cite{Kuznetsov2020} &  & \ce{OH -> O^*+H^+} \\
\hline
\ce{AgO} & 0.496 & 0.37\cite{Joya2016} &  & \ce{H2O+O^* -> OOH^*+H^+} \\
\hline
\ce{Bi4O7} & 0.221 & 0.800\cite{Thorarinsdottir2022} & 0.353 & \ce{OH -> O^*+H^+} \\
\hline
\ce{FeCo9O20} & 0.412 & 0.408\cite{Li2015}, 0.412\cite{Si2017} &  & \ce{H2O+O^* -> OOH^*+H^+} \\
\hline
\ce{Ag2SeO4} & 0.395 & 0.192\cite{Bibi2024} &  & \ce{H2O+O^* -> OOH^*+H^+} \\
\hline
\ce{Bi3SbO7} & 0.398 &  &  & \ce{OH -> O^*+H^+} \\
\hline
\ce{CaBi4O9} & 0.368 &  &  & \ce{OH -> O^*+H^+} \\
\hline
\ce{Li(CoO2)8} & 0.459 & 0.430\cite{Jiang2022} &  & \ce{OH -> O^*+H^+} \\
\hline
\ce{LiSbO3} & 0.611 &  &  & \ce{H2O+O^* -> OOH^*+H^+} \\
\hline
\ce{CoAgO3} & 0.306 & 0.310\cite{Zan2022} &  & \ce{OH -> O^*+H^+} \\
\hline
\ce{AgSnO3} & 0.491 &  &  & \ce{OH -> O^*+H^+} \\
\hline
\ce{Na(CoO2)3} & 0.416 & 0.236\cite{Sun2021} &  & \ce{OH -> O^*+H^+} \\
\hline
\ce{Ce9YO20} & 0.705 &  &  & \ce{OH -> O^*+H^+} \\
\hline
\ce{Ce4SnO10} & 0.594 &  &  & \ce{H2O+O^* -> OOH^*+H^+} \\
\hline
\ce{Ag6Mo10O33} & 0.546 & $>$ 0.540\cite{ArslanHamat2020} &  & \ce{OH -> O^*+H^+} \\
\hline
\ce{CdGe2O5} & 0.423 &  &  & \ce{OH -> O^*+H^+} \\
\hline
\ce{CuMoO4} & 0.460 &  &  & \ce{OH -> O^*+H^+} \\
\hline
\ce{ReAgO4} & 0.488 &  &  & \ce{OH -> O^*+H^+} \\
\hline
\ce{CuMoO4} & 0.626 &  &  & \ce{OH -> O^*+H^+} \\
\hline
\ce{Ce2Mo4O15} & 0.463 &  &  & \ce{OH -> O^*+H^+} \\
\hline
\ce{Ca(CuO2)2} & 0.598 &  &  & \ce{H2O+O^* -> OOH^*+H^+} \\
\hline
\ce{Cu3Mo2O9} & 0.545 &  &  & \ce{OH -> O^*+H^+} \\
\hline
\ce{Ti2CoO5} & 0.577 & 0.66\cite{Singh1990} &  & \ce{OH -> O^*+H^+} \\
\hline
\hline
\end{tabular}
\end{table}

\begin{table}[ht!]
\caption{\label{tab:best_facets_exp_comp_2}  Overpotentials from \gls{OC22}, the experimental literature, and \gls{DFT} (see Figure 7(b) in the main manuscript) along with the \gls{PDS} for our final set of 190 candidate (continued).}

\begin{tabular}{c|c|c|c|c}
\hline
\hline
 & $\eta$ (V) &  $\eta$ (V) &  $\eta$ (V) &  \\
Formula & (OC22) & (exp.) & (DFT) & PDS \\
\hline
\hline
\ce{TlCoO3} & 0.407 &  &  & \ce{H2O+O^* -> OOH^*+H^+} \\
\hline
\ce{Mn(SbO3)2} & 0.484 & 0.340\cite{Luke2021} &  & \ce{OOH^* -> O2+H^+} \\
\hline
\ce{CuWO4} & 0.391 & 0.270\cite{D1NJ04617A} &  & \ce{OH -> O^*+H^+} \\
\hline
\ce{NiBiO3} & 0.504 & 0.300\cite{Yu2022} &  & \ce{H2O+O^* -> OOH^*+H^+} \\
\hline
\ce{Fe(Bi5O8)5} & 0.699 & 0.420\cite{Arora2024} &  & \ce{H2O+O^* -> OOH^*+H^+} \\
\hline
\ce{Cu3(SbO3)4} & 0.366 &  &  & \ce{OH -> O^*+H^+} \\
\hline
\ce{Bi2O3} & 0.524 & 0.800\cite{Thorarinsdottir2022} & 0.292 & \ce{OH -> O^*+H^+} \\
\hline
\ce{Cr2Ag2O7} & 0.522 & $>$ 0.540\cite{ArslanHamat2020} &  & \ce{H2O+O^* -> OOH^*+H^+} \\
\hline
\ce{Cr2Ag2O7} & 0.426 & $>$ 0.540\cite{ArslanHamat2020} &  & \ce{H2O+O^* -> OOH^*+H^+} \\
\hline
\ce{AlTlO3} & 0.553 &  &  & \ce{OH -> O^*+H^+} \\
\hline
\ce{LiAgO2} & 0.388 &  &  & \ce{OH -> O^*+H^+} \\
\hline
\ce{Nb2Cu3O8} & 0.363 &  &  & \ce{OH -> O^*+H^+} \\
\hline
\ce{Cd(CoO2)2} & 0.431 &  &  & \ce{OH -> O^*+H^+} \\
\hline
\ce{Bi2O3} & 0.396 & 0.800\cite{Thorarinsdottir2022} &  & \ce{H2O+O^* -> OOH^*+H^+} \\
\hline
\ce{Ag2BiO3} & 0.343 & 0.700\cite{Simondson2022} &  & \ce{OH -> O^*+H^+} \\
\hline
\ce{K(CoO2)2} & 0.510 &  &  & \ce{OOH^* -> O2+H^+} \\
\hline
\ce{ZnCoO3} & 0.488 & $<$ 0.400\cite{Menezes2016}, 0.390-0.480\cite{Kim2014} &  & \ce{OH -> O^*+H^+} \\
\hline
\ce{Li(Bi3O5)4} & 0.473 &  &  & \ce{OOH^* -> O2+H^+} \\
\hline
\ce{CoAgO2} & 0.472 & 0.310\cite{Zan2022} &  & \ce{OH -> O^*+H^+} \\
\hline
\ce{Cu(BiO2)2} & 0.406 & 0.530\cite{Du2018} &  & \ce{H2O+O^* -> OOH^*+H^+} \\
\hline
\ce{CdCoO3} & 0.459 &  &  & \ce{OH -> O^*+H^+} \\
\hline
\ce{Cd2PbO4} & 0.220 &  & 0.222 & \ce{OH -> O^*+H^+} \\
\hline
\ce{Ag2GeO3} & 0.616 &  &  & \ce{OOH^* -> O2+H^+} \\
\hline
\ce{MnMoO5} & 0.564 & 0.570\cite{Balaghi2020} &  & \ce{OH -> O^*+H^+} \\
\hline
\ce{ScCoO3} & 0.341 &  &  & \ce{OH -> O^*+H^+} \\
\hline
\ce{Mn(SeO3)2} & 0.349 &  &  & \ce{OH -> O^*+H^+} \\
\hline
\ce{Cr3AgO8} & 0.620 & $>$ 0.540\cite{ArslanHamat2020} &  & \ce{H2O+O^* -> OOH^*+H^+} \\
\hline
\ce{InCoO3} & 0.582 & 0.370\cite{Hausmann2024} &  & \ce{OH -> O^*+H^+} \\
\hline
\ce{Ca(FeO2)2} & 0.586 & 0.320\cite{Lima2021} &  & \ce{OH -> O^*+H^+} \\
\hline
\ce{MnO2} & 0.406 & $>$ 0.600\cite{Hirai2016} &  & \ce{OOH^* -> O2+H^+} \\
\hline
\ce{TiAg2O3} & 0.468 & 0.650\cite{Bagheri2015} &  & \ce{H2O+O^* -> OOH^*+H^+} \\
\hline
\ce{CoPbO3} & 0.356 & 0.560\cite{Simondson2021} &  & \ce{OH -> O^*+H^+} \\
\hline
\ce{CaBiO3} & 0.352 &  &  & \ce{OH -> O^*+H^+} \\
\hline
\ce{Tl3Co3O8} & 0.503 &  &  & \ce{H2O+O^* -> OOH^*+H^+} \\
\hline
\ce{SrBiO3} & 0.392 &  &  & \ce{OH -> O^*+H^+} \\
\hline
\ce{Ce11O20} & 0.324 & 0.370\cite{Li2022} &  & \ce{OOH^* -> O2+H^+} \\
\hline
\hline
\end{tabular}
\end{table}

\begin{table}[ht!]
\caption{\label{tab:best_facets_exp_comp_3}  Overpotentials from \gls{OC22}, the experimental literature, and \gls{DFT} (see Figure 7(b) in the main manuscript) along with the \gls{PDS} for our final set of 190 candidate (continued).}

\begin{tabular}{c|c|c|c|c}
\hline
\hline
 & $\eta$ (V) &  $\eta$ (V) &  $\eta$ (V) &  \\
Formula & (OC22) & (exp.) & (DFT) & PDS \\
\hline
\hline
\ce{NaBi5O8} & 0.517 &  & 1.197 & \ce{OOH^* -> O2+H^+} \\
\hline
\ce{TlBiO4} & 0.333 &  &  & \ce{OOH^* -> O2+H^+} \\
\hline
\ce{Cu4Se3O10} & 0.520 & 0.440\cite{Tabassum2022} &  & \ce{OH -> O^*+H^+} \\
\hline
\ce{BaBiO3} & 0.631 &  &  & \ce{OOH^* -> O2+H^+} \\
\hline
\ce{Ce2(GeO3)3} & 0.546 &  &  & \ce{OH -> O^*+H^+} \\
\hline
\ce{CdSe2O5} & 0.659 &  &  & \ce{OH -> O^*+H^+} \\
\hline
\ce{AgBiO2} & 0.548 & 0.700\cite{Simondson2022} &  & \ce{OH -> O^*+H^+} \\
\hline
\ce{Zn(BiO2)2} & 0.429 &  &  & \ce{OH -> O^*+H^+} \\
\hline
\ce{CuSeO3} & 0.406 & 0.440\cite{Tabassum2022} &  & \ce{OH -> O^*+H^+} \\
\hline
\ce{Cu2SeO4} & 0.381 & 0.440\cite{Tabassum2022} &  & \ce{H2O+O^* -> OOH^*+H^+} \\
\hline
\ce{CdSeO3} & 0.488 &  &  & \ce{OOH^* -> O2+H^+} \\
\hline
\ce{FeSnO3} & 0.512 &  &  & \ce{OH -> O^*+H^+} \\
\hline
\ce{CoGeO3} & 0.496 & 0.340\cite{Xu2018} &  & \ce{OH -> O^*+H^+} \\
\hline
\ce{Hg2MoO4} & 0.259 &  &  & \ce{OH -> O^*+H^+} \\
\hline
\ce{CdIn2O4} & 0.457 &  &  & \ce{H2O+O^* -> OOH^*+H^+} \\
\hline
\ce{BaTl2O4} & 0.544 &  &  & \ce{H2O+O^* -> OOH^*+H^+} \\
\hline
\ce{Cr(SbO3)2} & 0.386 &  &  & \ce{H2O+O^* -> OOH^*+H^+} \\
\hline
\ce{Li3BiO4} & 0.414 &  &  & \ce{OH -> O^*+H^+} \\
\hline
\ce{SrSe2O5} & 0.254 &  &  & \ce{H2O+O^* -> OOH^*+H^+} \\
\hline
\ce{Ca2Se3O8} & 0.701 &  &  & \ce{OOH^* -> O2+H^+} \\
\hline
\ce{Cd6(CoO3)5} & 0.391 &  &  & \ce{H2O+O^* -> OOH^*+H^+} \\
\hline
\ce{Ag2SeO3} & 0.288 & 0.192\cite{Bibi2024} & 0.806 & \ce{H2O+O^* -> OOH^*+H^+} \\
\hline
\ce{BaSe2O5} & 0.661 &  &  & \ce{H2O+O^* -> OOH^*+H^+} \\
\hline
\ce{LuCoO3} & 0.559 &  &  & \ce{OOH^* -> O2+H^+} \\
\hline
\ce{Ag4GeO4} & 0.524 &  &  & \ce{OH -> O^*+H^+} \\
\hline
\ce{CuReO4} & 0.332 &  &  & \ce{OH -> O^*+H^+} \\
\hline
\ce{Co11CuO16} & 0.564 & 0.606\cite{Rajput2021} &  & \ce{OH -> O^*+H^+} \\
\hline
\ce{CoSe2O5} & 0.400 &  &  & \ce{H2O+O^* -> OOH^*+H^+} \\
\hline
\ce{Co5SbO8} & 0.486 & \cite{Luke2021} &  & \ce{OH -> O^*+H^+} \\
\hline
\ce{NaTlO2} & 0.489 &  &  & \ce{OH -> O^*+H^+} \\
\hline
\ce{MgIn2O4} & 0.417 &  &  & \ce{H2O+O^* -> OOH^*+H^+} \\
\hline
\ce{Li7Co5O12} & 0.381 & 0.430\cite{Jiang2022} &  & \ce{OH -> O^*+H^+} \\
\hline
\ce{Hg2WO4} & 0.679 &  &  & \ce{H2O+O^* -> OOH^*+H^+} \\
\hline
\ce{CuReO4} & 0.555 &  &  & \ce{OH -> O^*+H^+} \\
\hline
\ce{Sn2Ge2O7} & 0.552 &  &  & \ce{H2O+O^* -> OOH^*+H^+} \\
\hline
\ce{MnTlO3} & 0.196 &  & 0.080 & \ce{H2O+O^* -> OOH^*+H^+} \\
\hline
\hline
\end{tabular}
\end{table}

\begin{table}[ht!]
\caption{\label{tab:best_facets_exp_comp_4}  Overpotentials from \gls{OC22}, the experimental literature, and \gls{DFT} (see Figure 7(b) in the main manuscript) along with the \gls{PDS} for our final set of 190 candidate (continued).}

\begin{tabular}{c|c|c|c|c}
\hline
\hline
 & $\eta$ (V) &  $\eta$ (V) &  $\eta$ (V) &  \\
Formula & (OC22) & (exp.) & (DFT) & PDS \\
\hline
\hline
\ce{Ag2HgO2} & 0.473 &  &  & \ce{OH -> O^*+H^+} \\
\hline
\ce{Co2NiO4} & 0.504 & 0.390\cite{YuMingquan2022}, 0.316-0.438\cite{Cui2008} &  & \ce{OH -> O^*+H^+} \\
\hline
\ce{Ba8(Bi2O7)3} & 0.464 &  &  & \ce{OH -> O^*+H^+} \\
\hline
\ce{AgSbO4} & 0.319 &  &  & \ce{OOH^* -> O2+H^+} \\
\hline
\ce{MnTlO3} & 0.359 &  &  & \ce{OH -> O^*+H^+} \\
\hline
\ce{Ca(CoO2)2} & 0.321 & 0.331\cite{Lin2018} &  & \ce{H2O+O^* -> OOH^*+H^+} \\
\hline
\ce{Tl2SeO4} & 0.577 &  &  & \ce{OH -> O^*+H^+} \\
\hline
\ce{Tl2SeO4} & 0.520 &  &  & \ce{OH -> O^*+H^+} \\
\hline
\ce{CaSeO3} & 0.603 &  &  & \ce{OH -> O^*+H^+} \\
\hline
\ce{CdCu2O3} & 0.495 &  &  & \ce{OH -> O^*+H^+} \\
\hline
\ce{CuSbO4} & 0.364 &  &  & \ce{OH -> O^*+H^+} \\
\hline
\ce{YMn2O5} & 0.540 &  &  & \ce{OH -> O^*+H^+} \\
\hline
\ce{Cu2W2O7} & 0.425 & 0.270\cite{D1NJ04617A} &  & \ce{OH -> O^*+H^+} \\
\hline
\ce{Ca3WO6} & 0.488 &  &  & \ce{H2O -> OH*+H^+} \\
\hline
\ce{YCoO3} & 0.517 &  &  & \ce{OH -> O^*+H^+} \\
\hline
\ce{ZnCu2O3} & 0.423 & 0.550\cite{Yue2023} &  & \ce{OH -> O^*+H^+} \\
\hline
\ce{CuNiO2} & 0.401 & 0.580\cite{Wang2017} &  & \ce{H2O+O^* -> OOH^*+H^+} \\
\hline
\ce{MnBiO3} & 0.509 &  &  & \ce{OH -> O^*+H^+} \\
\hline
\hline
\end{tabular}
\end{table}

\begin{table}[ht!]
\caption{\label{tab:best_facets_exp_comp_5}  Overpotentials from \gls{OC22}, the experimental literature, and \gls{DFT} (see Figure 7(b) in the main manuscript) along with the \gls{PDS} for our final set of 190 candidate (continued). All materials listed from here are unstable as bulk materials ($E_{PBX} > 0.5$ eV) but can be stabilized as nanoparticles.}

\begin{tabular}{c|c|c|c|c}
\hline
\hline
 & $\eta$ (V) &  $\eta$ (V) &  $\eta$ (V) &  \\
Formula & (OC22) & (exp.) & (DFT) & PDS \\
\hline
\hline
\ce{Cr2WO6} & 0.252 &  &  & \ce{OOH^* -> O2+H^+} \\
\hline
\ce{TlCuO2} & 0.426 &  &  & \ce{OH -> O^*+H^+} \\
\hline
\ce{Y(FeO2)2} & 0.534 & 0.214\cite{Zhang2019b} &  & \ce{H2O+O^* -> OOH^*+H^+} \\
\hline
\ce{Sr2Tl2O5} & 0.374 &  &  & \ce{OOH^* -> O2+H^+} \\
\hline
\ce{ZrCoO3} & 0.458 & 0.400\cite{Huang2019} &  & \ce{OH -> O^*+H^+} \\
\hline
\ce{TiVO4} & 0.584 &  &  & \ce{H2O+O^* -> OOH^*+H^+} \\
\hline
\ce{HfFeO3} & 0.529 &  &  & \ce{OOH^* -> O2+H^+} \\
\hline
\ce{Mn4CuO8} & 0.386 & 0.150\cite{Zhou2019} &  & \ce{OH -> O^*+H^+} \\
\hline
\ce{CoCu2O3} & 0.417 & 0.606\cite{Rajput2021} &  & \ce{H2O+O^* -> OOH^*+H^+} \\
\hline
\ce{MnSe2O5} & 0.495 &  &  & \ce{OH -> O^*+H^+} \\
\hline
\ce{CrMoO4} & 0.416 &  &  & \ce{OH -> O^*+H^+} \\
\hline
\ce{KMn2O4} & 0.470 &  &  & \ce{H2O+O^* -> OOH^*+H^+} \\
\hline
\ce{Ba2Tl2O5} & 0.336 &  &  & \ce{OH -> O^*+H^+} \\
\hline
\ce{LuMnO3} & 0.544 &  &  & \ce{OH -> O^*+H^+} \\
\hline
\ce{TiMnO3} & 0.279 & 0.400\cite{Frydendal2015} & 0.559 & \ce{OH -> O^*+H^+} \\
\hline
\ce{KBiO2} & 0.267 &  & 0.278 & \ce{H2O -> OH*+H^+} \\
\hline
\ce{Na5ReO6} & 0.298 &  & 0.965 & \ce{OH -> O^*+H^+} \\
\hline
\ce{CuTeO4} & 0.537 &  &  & \ce{OH -> O^*+H^+} \\
\hline
\ce{ScCrO3} & 0.458 &  &  & \ce{H2O+O^* -> OOH^*+H^+} \\
\hline
\ce{Ta2CrO6} & 0.557 &  &  & \ce{OOH^* -> O2+H^+} \\
\hline
\ce{MnSnO3} & 0.428 &  &  & \ce{OH -> O^*+H^+} \\
\hline
\ce{Li4PbO4} & 0.661 &  &  & \ce{H2O -> OH*+H^+} \\
\hline
\ce{VSbO4} & 0.436 &  &  & \ce{OOH^* -> O2+H^+} \\
\hline
\ce{ScCuO2} & 0.407 &  &  & \ce{H2O+O^* -> OOH^*+H^+} \\
\hline
\ce{Mn2BeO4} & 0.324 &  &  & \ce{OOH^* -> O2+H^+} \\
\hline
\ce{AlCuO2} & 0.479 &  &  & \ce{H2O+O^* -> OOH^*+H^+} \\
\hline
\ce{TiCu3O4} & 0.375 &  &  & \ce{H2O+O^* -> OOH^*+H^+} \\
\hline
\ce{Ag2PbO2} & 0.434 &  &  & \ce{H2O+O^* -> OOH^*+H^+} \\
\hline
\ce{LuCrO3} & 0.435 &  &  & \ce{OH -> O^*+H^+} \\
\hline
\ce{VSeO4} & 0.458 &  &  & \ce{OH -> O^*+H^+} \\
\hline
\ce{Ag3RuO4} & 0.601 &  &  & \ce{H2O+O^* -> OOH^*+H^+} \\
\hline
\ce{GePb5O7} & 0.684 &  &  & \ce{OH -> O^*+H^+} \\
\hline
\ce{VCrO4} & 0.320 &  &  & \ce{OOH^* -> O2+H^+} \\
\hline
\ce{TlTeO4} & 0.496 &  &  & \ce{H2O -> OH*+H^+} \\
\hline
\ce{MnSeO3} & 0.438 &  &  & \ce{OH -> O^*+H^+} \\
\hline
\hline
\end{tabular}
\end{table}

\begin{table}[ht!]
\caption{\label{tab:best_facets_exp_comp_6}  Overpotentials from \gls{OC22}, the experimental literature, and \gls{DFT} (see Figure 7(b) in the main manuscript) along with the \gls{PDS} for our final set of 190 candidate (continued).}

\begin{tabular}{c|c|c|c|c}
\hline
\hline
 & $\eta$ (V) &  $\eta$ (V) &  $\eta$ (V) &  \\
Formula & (OC22) & (exp.) & (DFT) & PDS \\
\hline
\hline
\ce{Li3BiO3} & 0.523 &  &  & \ce{OH -> O^*+H^+} \\
\hline
\ce{VZn2O4} & 0.474 &  &  & \ce{H2O+O^* -> OOH^*+H^+} \\
\hline
\ce{Fe10O11} & 0.499 & 0.449\cite{Mullner2019} &  & \ce{H2O+O^* -> OOH^*+H^+} \\
\hline
\ce{Tl2SnO3} & 0.596 &  &  & \ce{H2O+O^* -> OOH^*+H^+} \\
\hline
\ce{ScMn2O4} & 0.285 &  & 0.328 & \ce{OH -> O^*+H^+} \\
\hline
\ce{ZrMnO3} & 0.501 &  &  & \ce{H2O+O^* -> OOH^*+H^+} \\
\hline
\ce{Li(CuO)2} & 0.515 &  &  & \ce{OOH^* -> O2+H^+} \\
\hline
\ce{Fe17O18} & 0.524 & 0.449\cite{Mullner2019} &  & \ce{OH -> O^*+H^+} \\
\hline
\ce{K6Co2O7} & 0.360 &  &  & \ce{OH -> O^*+H^+} \\
\hline
\ce{Cr2NiO4} & 0.381 & 0.334\cite{Babu2022} &  & \ce{H2O+O^* -> OOH^*+H^+} \\
\hline
\ce{YCuO2} & 0.380 &  &  & \ce{H2O+O^* -> OOH^*+H^+} \\
\hline
\ce{Mn2SnO4} & 0.425 &  &  & \ce{OH -> O^*+H^+} \\
\hline
\ce{MnCuO2} & 0.577 & 0.150\cite{Zhou2019} &  & \ce{H2O+O^* -> OOH^*+H^+} \\
\hline
\ce{YCuO2} & 0.377 &  &  & \ce{OH -> O^*+H^+} \\
\hline
\ce{Mn23FeO32} & 0.280 & 0.47\cite{Li2015} & 0.291 & \ce{OH -> O^*+H^+} \\
\hline
\ce{SrCr2O4} & 0.623 &  &  & \ce{H2O+O^* -> OOH^*+H^+} \\
\hline
\ce{KSnO2} & 0.194 &  & 1.314 & \ce{OH -> O^*+H^+} \\
\hline
\ce{CdRuO4} & 0.530 & 0.266\cite{Chen2020} &  & \ce{OOH^* -> O2+H^+} \\
\hline
\ce{Co(SbO2)2} & 0.660 & \cite{Luke2021} &  & \ce{OOH^* -> O2+H^+} \\
\hline
\ce{Mn2CrO4} & 0.371 & 0.367\cite{Song2016} &  & \ce{H2O+O^* -> OOH^*+H^+} \\
\hline
\ce{Na2Sb4O7} & 0.355 &  &  & \ce{OH -> O^*+H^+} \\
\hline
\ce{CeCrO3} & 0.293 &  &  & \ce{H2O+O^* -> OOH^*+H^+} \\
\hline
\ce{Na2Co2O3} & 0.297 & 0.236\cite{Sun2021} & 0.288 & \ce{OH -> O^*+H^+} \\
\hline
\ce{NaSb5O8} & 0.517 &  &  & \ce{H2O+O^* -> OOH^*+H^+} \\
\hline
\ce{RuPbO4} & 0.640 &  &  & \ce{OH -> O^*+H^+} \\
\hline
\ce{SnRuO4} & 0.321 &  &  & \ce{H2O+O^* -> OOH^*+H^+} \\
\hline
\ce{Ti(SnO2)2} & 0.476 &  &  & \ce{H2O+O^* -> OOH^*+H^+} \\
\hline
\ce{LiMn3O4} & 0.630 &  &  & \ce{OH -> O^*+H^+} \\
\hline
\ce{K2PbO2} & 0.549 &  &  & \ce{OOH^* -> O2+H^+} \\
\hline
\ce{Mn2NiO3} & 0.470 &  &  & \ce{OOH^* -> O2+H^+} \\
\hline
\ce{Mn3NiO4} & 0.541 &  &  & \ce{H2O+O^* -> OOH^*+H^+} \\
\hline
\ce{BaMn2O3} & 0.624 &  &  & \ce{OOH^* -> O2+H^+} \\
\hline
\ce{CaMn7O8} & 0.484 & 0.400-0.900\cite{Han2013} &  & \ce{OH -> O^*+H^+} \\
\hline
\hline
\end{tabular}
\end{table}

\clearpage

\end{document}